\documentclass{aa501}    %% AASTeX
\usepackage{natbib}
\usepackage{amssymb}
\usepackage{epsfig}
\begin{document}
\def\sizex{16.0 cm}
\def\bigx{10.0 cm}
\def\smallerxsize{7.0 cm}
\def\smallxsize{10.0 cm}
\def\smallysize{12.0 cm}

\title {The Canada-France deep fields survey-I\thanks{ Based on
    observations obtained at the Canada-France-Hawaii Telescope
    (CFHT) which is operated by the National Research Council of
    Canada, the Institut des Sciences de l'Univers (INSU) of the Centre
    National de la Recherche Scientifique and the University of Hawaii,
    and at the Cerro Tololo Inter-American Observatory and Mayall
    4-meter Telescopes, divisions of the National Optical Astronomy
    Observatories, which are operated by the Association of
    Universities for Research in Astronomy, Inc.  under cooperative
    agreement with the National Science Foundation.}}  \subtitle {
  100,000 galaxies, 1 deg$^2$: a precise measurement of
  $\omega(\theta)$ to $I_{AB}\sim25$}

\author{H.\ J.\ McCracken \inst{1} \and O. Le F\`evre \inst{1} \and M.\ 
  Brodwin \inst{2} \and S.\ Foucaud \inst{1} \and S.\ J.\ Lilly
  \inst{3} \and D. Crampton \inst {3} \and Y.\ Mellier \inst{4,5}}
\offprints {H.\ J.\ McCracken}
\date{received; accepted}

\institute{Laboratoire d'Astrophysique de Marseille, Traverse du
  Siphon, 13376 Marseille Cedex 12, France \and University of Toronto,
  Department of Astronomy, 60 St. George Street, Toronto, Ontario,
  Canada M5S 3H8 \and Herzberg Institute for Astrophysics, 5071 West
  Saanich Road, Victoria, British Colombia, Canada V9E 2E7 \and
  Institut d'Astrophysique de Paris, 98 bis Boulevard Arago, 75014
  Paris, France \and Observatoire de Paris, DEMIRM 61, Avenue de
  l'Observatoire, 75014 Paris, France}

\abstract{Using the University of Hawaii's 8K mosaic camera (UH8K), we
  have measured the angular correlation function $\omega(\theta)$ for
  100,000 galaxies distributed over four widely separated fields
  totalling $\sim 1\deg^2$ and reaching a limiting magnitude of
  $I_{AB}(3\sigma,3'')\sim 25.5$.  This unique combination of areal
  coverage and depth allows us to investigate the dependence of
  $\omega(\theta)$ at $1\arcmin$, $A_\omega(1\arcmin)$, on sample
  median magnitude in the range $19.5<I_{AB-med}<24$.  Furthermore, our
  rigorous control of systematic photometric and astrometric errors
  means that fainter than $I_{AB-med}\sim 22$ we measure
  $\omega(\theta)$ on scales of several arc-minutes to an accuracy of
  $30\%$. Our results show that $A_\omega(1\arcmin)$ decreases
  monotonically to $I_{AB}\sim25$. At bright magnitudes,
  $\omega(\theta)$ is consistent with a power-law of slope $\delta =
  -0.8$ for $0.2\arcmin<\theta<3.0\arcmin$ but at fainter magnitudes we
  detect a slope flattening with $\delta \sim -0.6$.  At the $3\sigma$
  level, our observations are still consistent with $\delta=-0.8$. We
  also find a clear dependence of $A_\omega(1\arcmin)$ on observed
  $(V-I)_{AB}$ colour.  In the magnitude ranges $18.5<I_{AB}<24.0$ and
  $18.5<I_{AB}<23.0$ we find galaxies with $2.6<(V-I)_{AB}<2.9$ (the
  reddest bin we consider) have $A_\omega(1\arcmin)$'s which are $\sim
  10\times$ higher than the full field population. On the basis of
  their similar colours and clustering properties, we tentatively
  identify these objects as a superset of the ``extremely red objects''
  found through optical-infrared selection. We demonstrate that our
  model predictions for the redshift distribution for the faint galaxy
  population are in good agreement with current spectroscopic
  observations. Using these predictions, we find that for low-$\Omega$
  cosmologies and assuming a local galaxy correlation length
  $r_0=4.3h^{-1}$~Mpc, in the range $19.5<I_{AB-med}<22$, the growth of
  galaxy clustering (parameterised by $\epsilon$), is $\epsilon\sim0$.
  However, at $22<I_{AB-med}<24.0$, our observations are consistent
  with $\epsilon\gtrsim 1$. Models with $\epsilon\sim 0$ cannot
  simultaneously match both bright and faint measurements of
  $A_\omega(1\arcmin)$. We show how this result is a natural
  consequence of the ``bias-free'' nature of the ``$\epsilon$''
  formalism and is consistent with the field galaxy population in the
  range $22.0<I_{AB}<24.0$ being dominated by galaxies of low intrinsic
  luminosity.   }

%\date{\fbox{\sc Draft Version: \today: DO NOT DISTRIBUTE!}}
\maketitle
\section{Introduction}
\label{sec:Int}

Understanding the formation of structure in the Universe is one of the
most pressing questions in modern cosmology. The Sloan and 2dF surveys
currently in progress \citep{1998wfsc.conf...77C,1995AAS...186.4405G}
will provide an accurate picture of large-scale structures in the local
Universe but presently our knowledge of galaxy clustering at $z>0.5$ is
poorly constrained. This is entirely a consequence of the technical
limitations in imaging and spectroscopic equipment, which (until
recently) have had fields of view $\sim 50$~arcmin$^2$; in all
cosmologies this translates to $<1~h^{-1}$Mpc at $z>0.5$. Covering a
substantial enough area to provide meaningful statistics on larger
($10-20 h^{-1}$Mpc) scales at higher redshifts ($z\sim1$) has been
prohibitively expensive in telescope time. Consequently, many galaxy
clustering measurements made at these redshifts have been dominated by
the effects of sample variance, and also have only been able to
investigate the highly non-linear regime where the predictions of
theoretical models for the clustering of galaxies are strongly
dependent on the biasing schemes employed.  Other studies, such as
investigating the variation of clustering amplitude by galaxy type or
intrinsic luminosity, or the accurate measurement of higher-order
statistics such as $S_3$ have been even more challenging.

However, with the advent of wide-field multi-object spectrographs like
VIRMOS and DEIMOS \citep{2000SPIE.4008..546L,1998wfsc.conf..333D}, in
addition to wide-field mosaic cameras on 4m-telescopes, this situation
is changing. In this paper we detail a new survey, the Canada-France
Deep Fields (CFDF) project which has been carried out using the
University of Hawaii's wide field ($28'\times28'$) 8K mosaic camera,
UH8K. This survey has targeted three of the original fields of the
Canada-France redshift survey \citep{1995ApJ...455...50L}. In total the
CFDF consists of four independent deep fields, each of area
$0.25$~deg$^2$. All of these have $VI$ colours, three $BVI$ and two and
a half $UBVI$.  The survey reaches a limiting magnitude ($3\sigma,3''$
aperture) of $I_{AB}\sim25.5$ and at least one magnitude fainter in
$UBV$ (Table~\ref{tab:cfdf.fields}).  The $\sim 10^5$ galaxies in the
survey, coupled with 1000 spectroscopic redshifts present throughout
our fields, allows us to investigate with unprecedented accuracy the
evolution of galaxy clustering to $z\sim1$ (the survey's median
redshift at its completeness limit of $I_{AB}\sim25.5$).  Moreover, our
four widely separated fields also ensure that we can estimate the
effect of cosmic variance on our results.

To date, there have been many studies of $\omega(\theta)$ carried out
using deep imaging surveys conducted using charge-coupled-device
(CCD)-based detectors
\citep{2000MNRAS.318..913M,1997ApJ...490...11W,1996ApJ...469..519H,BSM,RSMF}.
These works have generally focussed on one or two fields, usually
covering $\sim 50$~arcmin$^2$ each and typically reaching limiting
magnitudes of $I_{AB}\sim 25$. Several authors have also attempted to
cover larger areas ($>1$~deg$^2$) by mosaicing together many separate
pointings \citep{1999MNRAS.307..703R,1998ApJ...506...33P}, although
these surveys reach much shallower limiting magnitudes ($I_{AB} \sim
22$). In contrast, the CFDF survey, by virtue of its depth and angular
coverage, is able to provide an accurate measurement of
$\omega(\theta)$ in the range $18.5 < I_{AB} < 25.0$.

Normally the results from these surveys have been interpreted in terms
of the ``$\epsilon-$'' formalism first introduced to explain clustering
amplitudes observed at bright magnitudes on photographic plates
\citep{1977ApJ...217..385G,1978MNRAS.182..673P}. With this approach, an
assumed redshift distribution $dN/dz$ (or one measured from an
independent spectroscopic survey) and cosmology is coupled with a model
for the evolution of $\xi(r,z)$ (parametrised by $\epsilon$).  In this
way it is possible to predict the amplitude of $\omega(\theta)$ at any
magnitude limit, based on these assumptions. One can then conclude
which value of $\epsilon$ is most appropriate for any given set of
observations. Based on comparisons between $\omega(\theta)$
measurements in deeper CCD surveys and photographic measurements at
brighter magnitudes, many authors concluded that, for $z<1$ at least,
growth of galaxy clustering was consistent with $0<\epsilon<2$
\citep{BSM}. More recently, direct measurements of $r_0(z)$ have been
attempted at $z<1$ using spectroscopic samples
\citep{2000ApJ...542...57C,1999ApJ...524...31S,1996ApJ...461..534L,CEBC}.
These works have demonstrated the importance of sample selection in
measuring galaxy clustering evolution; $\epsilon$ has been shown to be
sensitive to the range of intrinsic galaxy luminosities and spectral
types selected. Attempts have also been made using photometric
redshifts computed using either ground-based or space-based imaging
data to measure the growth of clustering
\citep{2001ApJ...548..127T,2000ApJ...541..527B,1999MNRAS.310..540A,1998ApJ...499L.125C}.
However, the finding that clustering amplitudes for Lyman-break
galaxies was similar to some classes of galaxies found locally
\citep{1998ApJ...505...18A,1998ApJ...503..543G} has provided the
clearest evidence to date that this simple formalism could not fully
account for the observations of clustering at $z\sim3$.

In this paper, the first in a series, we will introduce the CFDF
survey, explain in detail our data reduction strategy and demonstrate
its robustness. As a first application of this dataset, we will present
a measurement of the projected galaxy correlation function
$\omega(\theta)$.  The angular size and depth of the CFDF allows us to
make a reliable determination of $\omega(\theta)$ over a large
magnitude range ($18 < I_{AB-med} < 24$).  Moreover the four separate
fields allows us to make an estimate of the field-to-field variance in
the galaxy clustering signal. Finally we will discuss how appropriate
the ``epsilon'' formalism is to describe the evolution of galaxy
clustering measured in our data.

In a future paper (Foucaud et al., in preparation) we will describe our
measurements of the clustering length $r_0$ at $z\sim3$ from a sample
of $\sim 2000$ Lyman-break galaxies derived from the CFDF dataset. By
adding $R-$ and $Z-$ band data from the new CFH12K camera
\citep{2000SPIE.4008.1022S} we expect to sufficiently increase the
accuracy of the photometric redshifts in the range $0<z<1$ to allow a
direct measurement of $r_0(z)$ in this interval; however, in this paper
we will concern ourselves only with measurement of $\omega(\theta)$ and
its dependence on apparent magnitude and colour.

\section{Observations and reductions}
\label{sec:observ-reduct}

\subsection{Observations}
\label{sec:observations-1}

$B$, $V$, $I$ observations were taken on the Canada-France-Hawaii
telescope with the UH8K mosaic camera \citep{1995AAS...187.7305M} over
a series of runs from December 96 -- June 97; details are given in
Table~\ref{tab:cfdf.fields}. Typically, for the $VI-$ band exposures we
used exposures of 1800s; for the $B-$ images we adopted exposure times
of 2400s. Individual exposures which had a FWHM $>1.2''$ were
discarded. Observing conditions were generally quite stable: for
example, for the 03hr field observations In $V$ and $I$ bands, the
median seeing is $\sim 1.1''$ and $1.2''$ respectively.  Additionally,
as our point-spread function (PSF) is almost always oversampled, it is
not necessary to carry out PSF homogenisation before image stacking. At
each pointing there is $\sim$ 10 exposures which allows us to carry out
adequate cosmic ray removal and to fill the gaps between each CCD in
the mosaic.

The UH8K camera consists of eight frontside-illuminated Loral-3
$2048\times4096$ CCDs, arranged in two banks of four devices each.
Each bank is read out sequentially. The upper-right CCD (number 8) has
very poor charge-transfer properties and data from this detector was
discarded.  The pixel scale is $0.205''$pixel$^{-1}$.  Additionally,
all of the CCDs have separate amplifier and controller electronics.
This arrangement, in addition to the necessity of removing the
CFHT-prime focus optical distortion before stacking our images, resulted
in a lengthy data reduction procedure which is outlined in the
following sections.

Because of the poor blue response of the Loral-3 devices, $U-$ band
observations for the CFDF survey were taken with the Kitt Peak 4m
Mayall telescope and the Cerro Tololo Inter-American Observatory's
(CTIO) 4.0m Blanco Telescope during a series of runs in 1997. Because
of the smaller field of view of these cameras, four separate pointings
were needed to cover each UH8K field. While the seeing on the $BVI$
frames is $\sim 0.7''-1.1''$ some of the $U-$ stacks are significantly
worse ($1.2''-1.4''$) which had to be accounted for during catalogue
preparation (two catalogues were prepared: one for those science
objectives which required $U-$ data and one for those which did not;
this is explained in more detail in Foucaud et al.).

\begin{table*}
\begin{tabular}{*{9}{c}}

{\bfseries Field}   & R.A. (2000) & Dec. (2000) & Band & Exposure time &  
Seeing  & $3\sigma$ measurement &   Area    & Date   \\
                    &             &             &      &    (hours)    & 
(arcsec) & ($AB$ mags)     & (deg$^2$) &        \\
\hline                                                                                                                       
                    &             &             &      &               &   
       &                            &           &        \\
{\bfseries 0300+00} & 03:02:40    & +00:10:21   & U    &     10.8      &   
1.0    &        26.98               &  0.25     & 06/97  \\
                    &             &             & B    &      5.5      &   
1.1    &        26.38               &  0.25     & 09/97  \\
                    &             &             & V    &      4.2      &   
1.3    &        26.40               &  0.25     & 12/96  \\
                    &             &             & I    &      5.5      &   
1.0    &        25.62               &  0.25     & 12/96  \\
                    &             &             &      &               &   
       &                            &           &        \\
{\bfseries 2215+00} & 22:17:48    & +00:17:13   & U    &     12.0      &   
1.4    &        27.56               &  0.12     & 06/97  \\
                    &             &             & B    &      5.5      &   
0.8    &        25.90               &  0.25     & 09/97  \\
                    &             &             & V    &      2.7      &   
1.0    &        26.31               &  0.25     & 06/97  \\
                    &             &             &      &      1.5      &   
       &                            &           & 09/97  \\
                    &             &             & I    &      1.7      &   
0.7    &        25.50               &  0.25     & 06/97  \\
                    &             &             &      &      2.7      &   
       &                            &           & 09/97  \\
                    &             &             &      &               &   
       &                            &           &        \\
{\bfseries 1415+52} & 14:17:54    & +52:30:31   & U    &     10.0      &   
1.4    &         27.71              &  0.25     & 03/97  \\
                    &             &             & B    &      5.3      &   
0.8    &         26.23              &  0.25     & 06/97  \\
                    &             &             & V    &      2.3      &   
1.0    &         25.98              &  0.25     & 06/97  \\
                    &             &             & I    &      2.6      &   
0.7    &         25.16              &  0.25     & 06/97  \\
                    &             &             &      &               &   
       &                            &           &        \\
{\bfseries 1130+00} & 11:30:02    & -00:00:05   & V    &      3.3      &   
0.8    &         26.42              &  0.25     & 05/97  \\
                    &             &             & I    &      4.4      &   
1.0    &         25.80              &  0.25     & 12/96  \\
                    &             &             &      &      3.0      &   
       &                            &           & 05/97  \\
                    &             &             &      &               &   
       &                            &           &        \\
\end{tabular}
\caption {Details of the images used in the CFDF fields. For each field
  we list the total integration time, in addition to the $3\sigma$
  detection limit inside an aperture of $3''$. For the 22hr and 11hr
  fields, the fields are composed of two separate stacks with bonette
  rotations, as detailed in the text. For the 22hr field the $U-$
  data covers only half the field.}
\label{tab:cfdf.fields}
\end{table*}

\subsection{Preprocessing}
\label{sec:prepr-flat-field}

Pre-processing followed the normal steps of overscan correction, bias
subtraction and dark subtraction. We fit a fourth-order Legendre
polynomial to the overscan region to allow us to remove structure in
the overscan pattern. The high dark current ($\sim 0.1 e^-$s$^{-1}$) of
the UH8K makes it essential to take dark frames.  For $I-$ and $V-$
bandpasses, where the sky background is high, one unique dark frame
(composed of the average of 5-10 individual exposures) can be used;
however, with the $B-$ band special steps have to be taken due to the
``'dark-current jumping'' effect which manifests itself as a either
``high'' or ''low'' dark current level, which can affect either right
or left banks of CCDs independently. Because of the low quantum
efficiency of the UH8K CCDs in $B$ the dark current is a significant
fraction of the sky level and consequently accurate dark subtraction
\emph{must} be performed to produce acceptable results. We achieve this
by generating two sets of darks: ``high'' darks and ``low'' darks which
we apply by a trial-and-error method to each $B-$exposure to determine
which dark is appropriate for a given dataset.  The high and low darks
are identified by their statistics (mean, median).  Two full reductions
are done.  Each kind of dark is subtracted and then a dark-independent
flat (dome or twilight) is applied.  The flatter final image (with
smallest amplitude of residual flatness variations) indicates the kind
of dark that was actually present in the data.

We generate ``superflats'' from the science images themselves as dome
flats or twilight flats by themselves produce residual sky variations
$>1\%$. These superflats are constructed by an iterative process,
which begins by the division of our images by a twilight flat. On
these twilight-flattened images we run the \texttt{sextractor}
\citep{1996A&AS..117..393B} package to produce mask files which
identify the bright objects on each frame. For large saturated stellar
objects we grow these masks well into the wings of the
point-spread-function by placing down circles on these objects. In
addition, we mask out non-circular transients (typically scattered
light and saturated columns near bright stars) on some images. Using
these mask files we combine each image to produce a superflat each
pixel of which contains only contributions from the sky and not object
pixels. After the division of the twilight-flattened images by the
superflats, the residual variation in the sky level is $<1\%$.

Because the gain and response of each CCD in the mosaic is not the
same, we scale our flat fields for each filter so that the sky
background in each chip after division by the flat field is the same
(normally these exposures are scaled to chip 7). In
Sect.~\ref{sec:syst-photm-errors-1} we will quantify how successful
this procedure is in restoring a uniform zero-point over the entire
field of view of the image.

\subsection{Astrometric image mapping}
\label{ast.image}

For each set of observations in each filter of our field, we have
typically $\sim 10$ pointings (we use the term ``pointing'' to refer to
the eight separate images which comprise each read-out of the UH8K
camera).  Each of these pointings are offset by $\sim 5 ''$ from the
previous one in a random manner; these offsets allows us to remove
transient events and cosmetic defects from the final stacks, and also
to ensure that the gaps between the CCDs (which are $\sim 3''$) are
fully sampled .  Additionally, on some of our fields, pointings were
taken over several runs with the camera bonette in different
orientations. Given the non-negligible optical distortion at the CFHT
prime focus (amounting to a displacement of several pixels at the edge
of the field relative to an uniform pixel scale) this means it is
essential that these distortions are removed before the pointings can
be coadded to produce a final stack. A further requirement is that each
of the stacks for each of the filters can also be accurately
co-aligned, for the purposes of measuring aperture colours reliably.

In the mapping process, the images from each CCD are projected onto an
undistorted, uniform pixel plane. The tangent point in this plane is
defined as the optical centre of the camera, and is the same for each
of the eight CCDs.  Overall, our goal is to produce an root-mean-square
registration error between pointings in each dither sequence and
between stacks constructed in different filters which does not exceed
one pixel (0.205'') over the entire field of view.

Our astrometric mapping process is essentially a two-step process. In
outline, this involves first using the United States Naval Observatory
(USNO) catalogue \citep{1998AAS...19312003M} to derive an absolute
transformation between (x,y) (pixels) to $(\alpha,\delta)$ (celestial
co-ordinates).  Following this, a second solution is computed using
sources within each field. This method ensures that our pointings are
tied to an external reference frame, but also provides sufficient
accuracy to ensure that pointings can be registered with the precision
we require; the surface density and positional accuracy of the USNO
stars is too low to ensure this. To fully characterise the distortions
which are present in the camera optics we adopt a higher
order-solution, which consists of a combination of a standard tangent
plane projection and higher-order polynomial terms. To prevent solution
instabilities at the detector edges we use a third-order polynomial
solution. To compute the astrometric solutions and carry out the image
mapping we use the {\tt mscred} package provided within the
IRAF\footnote{IRAF is distributed by the National Optical Astronomy
  Observatories, which are operated by the Association of Universities
  for Research in Astronomy, Inc., under cooperative agreement with the
  National Science Foundation.} data reduction environment.

For each field we begin our procedure with the $I-$ band, as these
exposures normally have the highest numbers of objects. Using the
external catalogue we compute an astrometric solution containing a
common tangent point for each of the eight CCDs. Typically, we find ~50
-- 100 sources per CCD, with a fit RMS of $\sim 0.3''$. Next, using the
task {\tt mscimage} we project each of the eight images onto the
undistorted tangent plane, using a third-order polynomial
interpolation. Following this, we extract the positions in celestial
co-ordinates of a large number ($\sim 1000$) of sources distributed
over all eight CCDs. This list forms our co-ordinate reference, and we
use this list in conjunction with the task {\tt mscimatch} to correct
for the adjustments in the WCS (world co-ordinate system) due to slight
rotation and scale change effects for each successive pointing in the
dither. Before stacking we also remove residual gradients by fitting a
linear surface and scale the images to photometric observations if
necessary. Because each image now has a uniform pixel scale we need
only apply linear offsets before constructing the final stack. Setting
the gaps between each CCD to large negative values which are rejected
in the stacking process allows the production of a final, contiguous
image. To stack our images we use a clipped median, which although not
optimal in signal-to-noise terms, provides the best rejection of
outlying pixels for small numbers of pointings.

From this final, combined stack, we extract a second catalogue of
($\sim 1000$) sources (with $\alpha$,$\delta$ computed from the
astrometric solution) which we use as an input 'astrometric catalogue'
for the dither sequences observed in other filters (as opposed to the
USNO catalogues which we use for the first step). Typically, the RMS of
the fit these cases is $<0.1''$.  We then proceed as before, mapping
each of CCD images from each pointing in the dither set onto the
undistorted tangent plane and constructing a final stack.

\begin{figure}
\resizebox{\hsize}{!}{\includegraphics{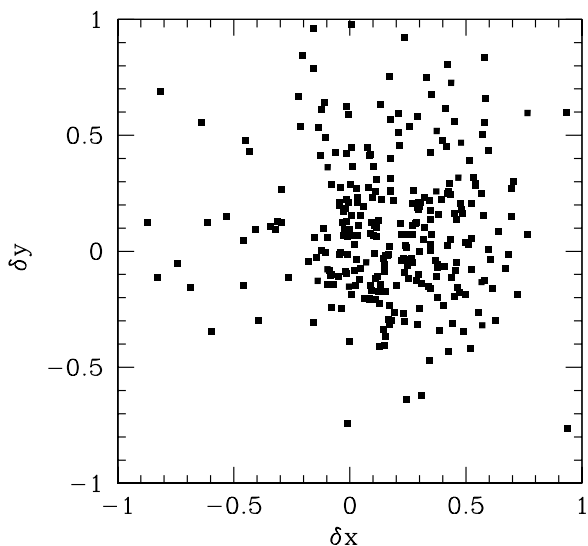}}
  \caption{The difference, in pixels, of the positions of non-saturated
    stellar sources with $18 < I < 22$ in the 03hr $I-$ band stack
    compared with the 03hr $B-$ stack. The rms in both co-ordinate
    directions is $\sim 0.3$ pixels, or $\sim 0.06''$.}
  \label{fig:btoiastro}
\end{figure}

For the final mapping between the stacks taken in different filters, we
find a residual of $\sim 0.06''$, or $\sim 0.3$ pixels over the whole
field of view, which is with our aim of a root-mean-square of one pixel
or less; this is illustrated in Fig. \ref{fig:btoiastro}.  Our large
grid of reference stars extracted from the $I-$ band image ensures that
the derived WCS for the other filters is very well matched to the $I-$
band exposure. We also find that this method allows us to successfully
register and combine observations distributed over separate runs
containing bonette rotations.

For the $U-$ band exposures, each of the four corners were stacked
separately and scaled to have the same photometric zero-point. Then,
using the $I-$ band reference list described previously, a mapping was
computed between each of the corners and the undistorted $I-$ stack.
During this process the image was also resampled (using the same
third-order polynomial kernel employed above) to have the same pixel
scale as the UH8K data. These four $U-$ images were then stacked to
produce the final $U-$ mosaic.  Overall, we find that the rms of the
mapping between $I-$ and $U-$ is not as good as between the UH8K data,
with an rms $\sim 1$ pixel in the region of CCD 8 (the CCD suffering
cosmetic defects) but still within our stated goal.

Each of the final images have a scale of 0.204''/pixel and cover $\sim
8000 \times 8000$ pixels (the scale of the final stack is determined
from the linear part of the astrometric solution of the image which is
closest to the tangent point of the camera). In all the analyses that
follow we exclude the region covered by CCD 8 as this chip has very bad
charge transfer properties and is highly photometrically non-linear.
However, for the 11hr-I and 22hr-I stacks we are able to use the full
area of UH8K because these final stacks consist of two separate stacks
with bonette rotation, allowing us to cover the region lost by the bad
CCD.

\subsection{Calibration}
\label{sec:phot-reduct}
\begin{table}
\begin{tabular}{*{7}{c}}

{\bfseries Field}   & $b$   & $l$   & $E(B-V)^a$   & $E(B-V)^b$               \\
                    &     &     & Schlegel et al.& BH                 \\
\hline               
                    &     &     &          &                      \\
{\bfseries 0300+00} & -48 & 179 & 0.071  & 0.040              \\
                    &     &     &          &                      \\
{\bfseries 2215+00} & -44 & 63  & 0.061  & 0.040              \\
                    &     &     &          &                      \\
{\bfseries 1415+52} & +60 & 97  & 0.011  & 0.000                    \\
                    &     &     &          &                      \\
{\bfseries 1130+00} & +57 & 264 & 0.026  & 0.013              \\
                    &     &     &          &                      \\
\end{tabular}
\caption {The $(l,b)$ for each CFDF field, together with the galactic
  dust extinction corrections from \citet{1998ApJ...500..525S}($^a$)
   compared to the values of \citet{1982AJ.....87.1165B} ($^b$, BH).}
\label{tab:schlegel}
\end{table}

In this Section we will describe how we derive the relationship between
magnitudes measured in our detector/filter combination (which we denote
by $u_{cfdf}, b_{cfdf}, v_{cfdf}, i_{cfdf}$) and the standard Johnson
UBVI system. Our zero-points are computed from observations of the
standard star fields of \citet{1992AJ....104..340L}. Of the
four observing runs with UH8K which are discussed here, only the data
from May were not photometric and for the runs of June and October
sufficiently large numbers of observations of standards were taken
($n\sim 30$) it was possible to determine the colour equation.

We apply the same data reduction procedure to the standard star fields
as we do to the science frames. This involves bias and dark
subtraction, followed by flat-fielding which is necessary to account
for the sensitivity and gain variations from CCD to CCD and the
application of the astrometric solution derived previously to produce a
single, undistorted image. This procedure assures that a uniform pixel
scale is restored when computing the photometric zero-point, and has
added advantage of that we may use a catalogue of standards in
$\alpha,\delta$ to derive the zero-point in a semi-automated fashion.
All standards used were visually inspected and faint or saturated
objects were rejected. Our zero-points are corrected for galactic
extinction using the $E(B-V)$ values provided by
\citet{1998ApJ...500..525S}.

\begin{figure}
\resizebox{\hsize}{!}{\includegraphics{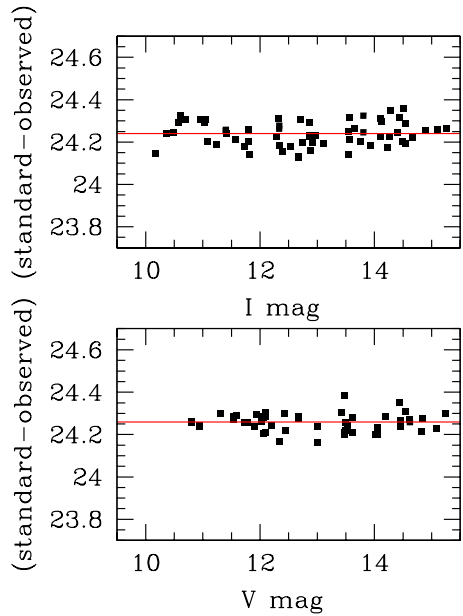}}
  \caption{Standard star calibration plots for the $I$ and $V$ filters
    for the October 97 data (upper and lower panels respectively). For
    each panel, we plot the difference between the standard star
    magnitude and the instrumental magnitude (corrected to one airmass)
    as a function of the true $I,V$ magnitudes. The solid line shows
    the adopted zero-point; this is $24.24\pm 0.04 $ ($V$) and
    $24.26\pm 0.06$ ($I$), per second and at one airmass.  Atmospheric
    extinction coefficients are taken from the CFHT handbook, although
    most of our standards are at or near the zenith.}
  \label{fig:viplot}
\end{figure}

In Fig.~\ref{fig:viplot} we show sample plots of standard star
observations taken during the October observations. For the $I$ and $V$
band for all three runs we derive zero-point rms of $\sim 0.05$
magnitudes, with no evidence for position-dependent residuals (as would
happen if an error had occurred in the flat-fielding process).

Our observations do not indicate that the presence of a colour term for
either the $V$ or $I$ filters and in what follows we assume
$V=v_{cfdf}$ and $I$=$i_{cfdf}$. However, we find that $b_{cfdf}$ is
different from the standard Johnson $B-$ and for this reason we derive
colour terms.

\begin{equation}
b_{cfdf}\sim B-0.07(B-V)-23.2 .
\label{eq:bveq}
\end{equation}

We convert our magnitudes to the $AB$ system using
$U_{AB}=u_{cfdf}+0.73$, $B_{AB}=b_{cfdf}-0.1$, $V_{AB}=v_{cfdf}$ and 
$I_{AB}=i_{cfdf}+0.43$ (computed based on our filter response
functions). 

\begin{figure}
\resizebox{\hsize}{!}{\includegraphics{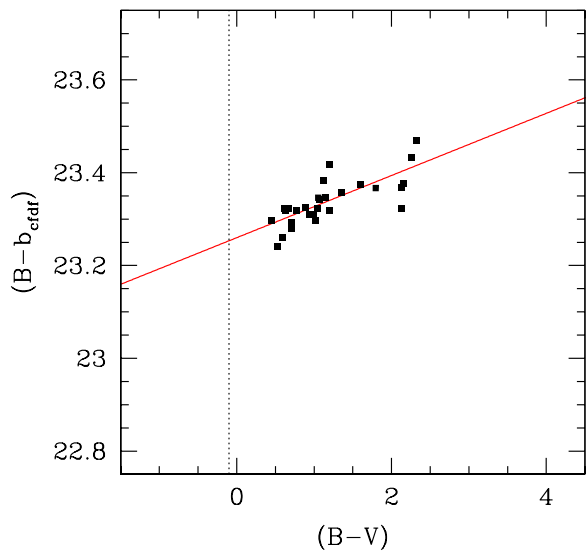}}
  \caption{$B-b_{cfdf}$ \textit{vs} 
    $(B-V)$, based on observations  of $51$ standard
    stars during the October 1997 observations. The slope and offset of 
    the fitted line is 0.07 and 23.26, respectively. The slope is not
    well determined due to the small range in (B-V) colours spanned by
    our observations and we expect errors of $\sim 0.02$.}
  \label{fig:bvplot}
\end{figure}

\subsection{Systematic photometric errors}
\label{sec:syst-photm-errors-1}

To measure accurately the galaxy clustering signal it is essential that
the photometric zero-point is uniform across the stacked images.
Zero-point variation across the mosaic will introduce excess power on
large scales and contribute to a flattening of the $\omega(\theta)$ on
large scales. For single-CCD images, improperly flattened data can
produce this effect; in our case we have the additional complication
that we must correctly account for the different responses and
amplifier gains for the eight CCDs in our mosaic. As outlined above,
this is accomplished by scaling each CCD image before co-addition to
have the same sky background. Our standard star reductions detailed in
Sect.~\ref{sec:phot-reduct} have already indicated that this
procedure produces zero-point variations on order $0.05$ magnitudes
r.m.s. . However, further observations allow a more rigorous test of
this effect. On two separate occasions we have observed the same field
(11hr, 22hr) after the camera bonnette had been rotated $180\deg$.
These data provide an excellent opportunity to verify that there are no
residual systematic magnitude zeropoint variations in our final stacks
after co-addition and stacking have been carried out.

To carry out these tests we prepare two separate stacked mosaics. For
the field at 11hrs, we have 4.4hrs of integration in I from December
1996 and 3hrs total integration from May 1997. By using sources
extracted from the December run to compute our astrometric solution
following methods outlined above, we can produce final stacked mosaics
which are aligned with a standard deviation of $< 1$ pixel over the
entire field of view.  By carrying out the detection process on the sum
of these two images, and photometry on the two separate mosaics, we are
able to measure the difference in magnitude between sources located in
the same part of the sky but falling on different elements of the
detector-telescope system. Note that because we place our photometric
apertures on the same positions on each of the two stacks, this test
also allows us to investigate magnitude errors introduced by mapping
inaccuracies between the two stacks, which are expected to be present
for the measurement of aperture colours.

\begin{figure}
\resizebox{\hsize}{!}{\includegraphics{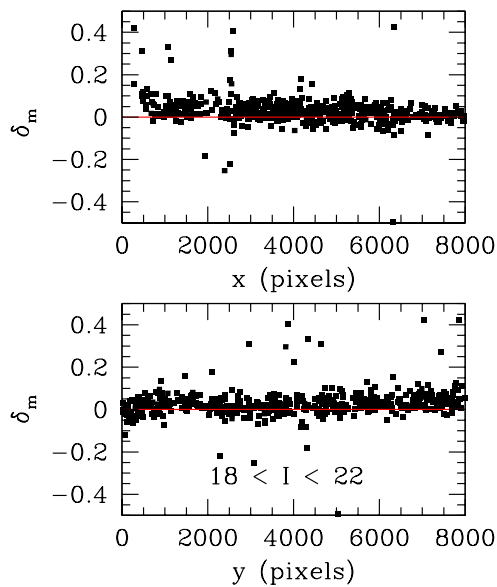}}
  \caption{The magnitude difference as a function of $x$ and $y$
    position between non-saturated stellar sources with $18.5 <
    I_{AB}<22.5$ in the December 1996 11hr stack and the May 1997 data
    covering the same field. We find $\sigma (\delta_m)\sim 0.04$
    magnitudes. }
  \label{fig:syserrors1}
\end{figure}

The results of this test are illustrated in Fig. \ref{fig:syserrors1}
where we plot the difference in magnitude for non-saturated stellar
sources with $18.5 < I_{AB} < 22.5$ between the two 11hr stacks as a
function of position in both $x$ and $y$ directions. We find that the
systematic magnitude errors, measured as the dispersion of these
residuals is $\sim 0.04$ magnitudes, which corresponds to the limit of
our CCD-to-CCD calibrations, as explained in
Sect.~\ref{sec:phot-reduct}.

Towards the $50\%$ completeness limit of our catalogues,
$I_{AB}\sim25.5$, differential incompleteness becomes a significant
bias in the measurement of $\omega(\theta)$. This effect arises from
the differing read-out electronics and detector gains used in each of
the eight individual CCDs in UH8K. \citet{1995ApJ...439...14N}, using
the Palomar four-shooter camera, discuss this effect in more detail.
However, we emphasise that all our scientific analysis is only carried
out where our completeness is $>80\%$, as determined from the
simulations and source counts detailed in the following sections.
Furthermore have verified that this effect is only significant at the
faintest magnitudes by adding 40,000 objects with the same clustering
amplitude as galaxies at $I_{AB}\sim 25$ to one of our images.  This
test is described in detail in Sect.~\ref{sec:magn-limit-sampl}.

\subsection {Random photometric errors}
\label{sec:syst-photm-errors}

We may also use these repeated observations of the same field to
investigate what \emph{random} photometric errors are present in our
data. At fainter magnitudes these errors dominate. We have used three
separate methods to estimate the magnitude errors computed in our data;
firstly, we may use the errors computed directly by
\texttt{sextractor}; secondly, we can use the errors computed from the
simulations detailed in Sect.~\ref{sec:incompl-simul} in which
stellar objects are added to our fields and recovered; and lastly, we
may use the our two independent stacks of the same field to estimate
our errors.

\begin{figure}
\resizebox{\hsize}{!}{\includegraphics{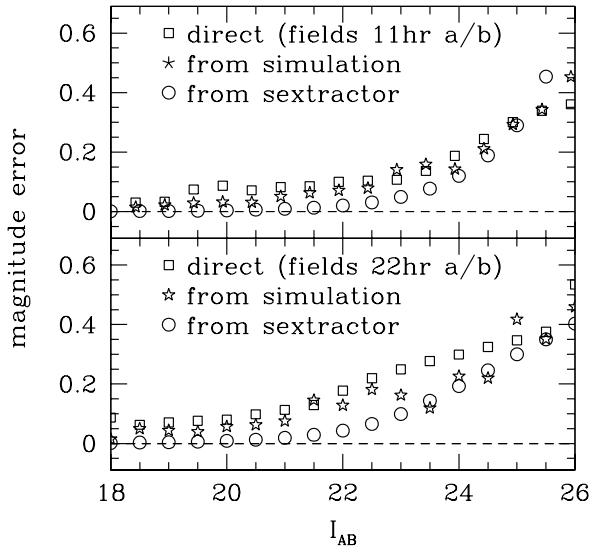}}
  \caption{Root-mean-square (RMS) magnitude errors as a function of $I_{AB}$
    magnitude. The upper and lower panels shows measurements made on
    the 11hr and 22hr fields respectively.  Magnitude errors were
    computed using a variety of techniques: from sextractor (open
    circles); from a simulation in which stars were added to random
    patches of the field (stars) and the dispersion in the recovered
    total magnitudes is calculated; and finally as measured between two
    independent stacks covering the same field (open squares).  The
    direct magnitude error measurements have been multiplied by $\sqrt
    2$ to account for the shorter exposure time for the individual
    stacks.  At all magnitudes the sextractor errors appear to
    underestimate the both the direct and simulation error by at least
    a factor of two.}
  \label{fig:phot-random}
\end{figure}

Fig.~\ref{fig:phot-random} shows magnitude errors for these three
different estimators: \texttt{sextractor} (circles), the simulation
(stars) and from the direct measurement (squares). We have carried out
these tests on both the 11hr stacks and the 22hr stacks. In many
magnitude ranges, the \texttt{sextractor} errors are lower than the
other two measurements. We believe the origin of this discrepancy is
due to the image resampling and interpolation process which produces
images with correlated background noise. By contrast, the
\texttt{sextractor} magnitude errors are computed assuming white
background noise.

\subsection{Incompleteness simulations and limiting magnitudes }
\label{sec:incompl-simul}

\begin{figure}
\resizebox{\hsize}{!}{\includegraphics{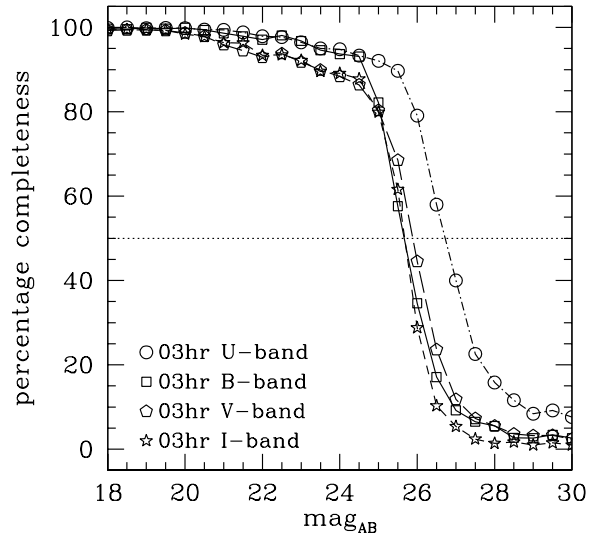}}
  \caption{Derived completeness in four bands for the 03hr field as a
    function of AB magnitudes. These simulations were derived by adding
    1,000 artificial stars in each half-magnitude interval to the
    original images and measuring how many objects were recovered as a
    function of magnitude. The detection threshold used for this test
    (number of sigmas above the noise background and minimum number of
    connected pixels) were the same as used for the actual detections
    on the $\chi^2$ image.}
  \label{fig:completeness}
\end{figure}

In Table~\ref{tab:cfdf.fields} we list the $3\sigma$ values for
detection in a 3'' aperture. These should be regarded as \textit{lower}
limits on the detectability of the galaxies in our catalogues. To
better characterise the photometric properties of our images we have
carried out an extensive set of simulations. These simulations involved
adding artificial stars and galaxies to our single-band images and
measuring the fraction recovered as a function of magnitude. In
Fig.~\ref{fig:completeness} we show the results of one set of such
simulations for the 03hr field.

We note that this result should be regarded as \emph{lower limit} to
the completeness in our data as our actual catalogues are constructed
using the chisquared technique described in
Sect.~\ref{sec:making-match-catal} and can be expected to be slightly
deeper (but note also that in constructing the chisquared catalogues
all images must first be convolved to the worst seeing). From a rough
comparison of the $I-$band galaxy counts presented in
Fig.~\ref{fig:countsIAB}, we see that the we see that the simulations
provide a good estimate of the magnitude at which the the observed
counts begin to fall off.

\subsection{Comparison with CFRS photometry}
\label{sec:comp-with-prev}

Three of our fields (22hr, 03hr, 14hr) cover the original survey fields
of the Canada-France Redshift Survey (CFRS;
\citet{1995ApJ...455...50L}). For the 22hr and 14hr field we have $BVI$
photometry from the CFRS; for the 03hr field, data exists in the $VI$
bandpasses.

\begin{figure}
\resizebox{\hsize}{!}{\includegraphics{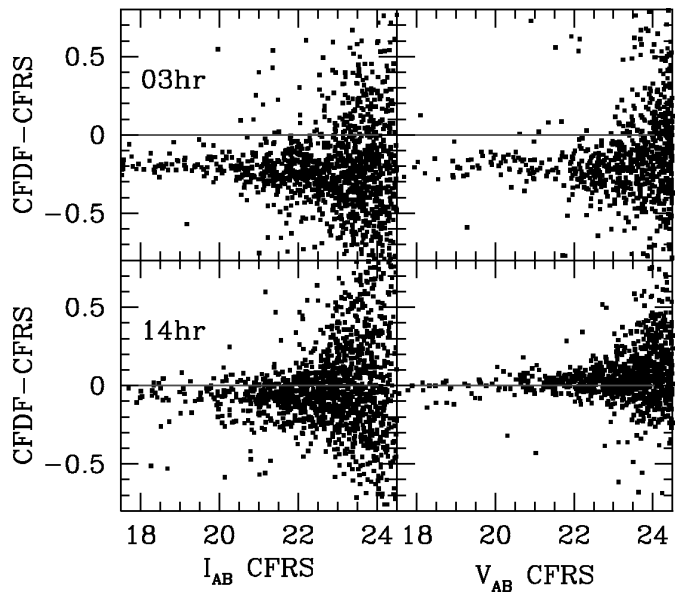}}
\hfill
  \caption{Comparison between CFDF and CFRS magnitudes measured in 3''
    aperture diameter for the $V$-band (right panels) and $I-$ band
    (left panels). The $\sim 0.2$ magnitude offset between CFRS and
    CFDF magnitudes for the 03hr field is a consequence of the
    significant galactic extinction in this field, which the CFRS
    magnitudes have not been corrected for.}

  \label{fig:cfdf_cfrs_VI}
\end{figure}

For all these fields we have carried out a detailed comparison of our
photometry with CFRS photometry. In Fig.~\ref{fig:cfdf_cfrs_VI} we
compare $V$ and $I$ photometry from our stacked images with the CFRS
for $V$ and $I$ filters in the 03hr and 14hr fields. For the 14hr
fields, the agreement with the CFRS photometry is $<0.1$ magnitudes or
better. In the 03hr field, however, we find that our magnitudes are
$\sim 0.2$ and $0.1$ magnitudes brighter than CFRS magnitudes in the
$V$ and $I$ filters respectively.  We suspect the origin of this
discrepancy is that the CFRS fields were selected to have low galactic
extinction as measured in the maps of \citet{1982AJ.....87.1165B} (BH).
In Table~\ref{tab:schlegel} we show that the difference between the BH
extinction and the more recent $E(B-V)$ values given in
\citet{1998ApJ...500..525S} is non-negligible (amounting to $\sim 0.15$
in $I_{AB}$ magnitudes). In all our fields we apply extinction
corrections based on $E(B-V)$ values from \citet{1998ApJ...500..525S}. 

\section{Catalogue preparation}
\label{sec:catal-prep}

\subsection{Construction of merged catalogues}
\label{sec:making-match-catal}

To prepare merged catalogues with $UBVI$ colours for each object we
derive a ``detection'' image to locate objects and then perform
aperture photometry on each separate filter at the positions defined by
the detection image. This procedure avoids the difficulties associated
with merging separate single-band detection images (such as differences
in object centroids between $U-$ and $I-$ band images, for example).
This detection image is constructed using a $\chi^2$ image technique
\citep{1999AJ....117...68S}, expressed in equation~(\ref{eq:1}), where
$a_i$ represents the background-subtracted pixel value in filter $i$,
$\sigma_i$ the rms noise at this pixel and $n$ is the number of filter,

\begin{equation}
  \chi^2 = {1\over n} \sum_{i=1}^{i=n}(a_i/\sigma_i)^2 .
\label{eq:1}
\end{equation}

This image has the advantage over other (more arbitrary) combinations
of images such as $(V+I)$ in that it has a simple physical
interpretation, namely that each pixel of this image represents the
probability of detecting an object at that location.  We compute $a_i$
and $\sigma_i$ for each pixel in each of the stacks using
\texttt{sextractor}; this procedure allows us to correctly account for
regions of varying signal-to-noise such as the overlap regions at the
CCD boundaries. This resulting image is then used as input to
\texttt{sextractor} as a detection image in the dual-image mode.  We
note that this method requires that both images are convolved to have
the same full-width at half maximum and furthermore that they have a
positional accuracy between filters of better than 1 pixel (as we have
demonstrated in Sect.~\ref{ast.image} our internal positional
accuracy is $\sim 0.3$ pixels, which meets this objective).
\nocite{astro-ph9812105}
We use an empirical approach to set the detection threshold in the
chisquared image, similar to that employed in da Costa et al. (1998).
Based on the numbers of objects detected in ``blank'' images (frames
which have the same background noise as our real images), the noise
threshold is lowered in the chisquared image until the the number of
additional sources detected is less than twice the number of sources
detected in the blank images for the same change in the threshold. We
emphasise however that the exact choice of the threshold is unimportant
in this work as the range of variations considered in this procedure
($\sim 2.0\sigma$) does not affect object detection even at the
faintest magnitude limit where we carry out our scientific analysis
($I_{AB}\sim25$).

\subsection{Effect of $\chi^2$-technique on galaxy magnitudes}
\label{sec:effect-chisq-techn}
\begin{figure}
\resizebox{\hsize}{!}{\includegraphics{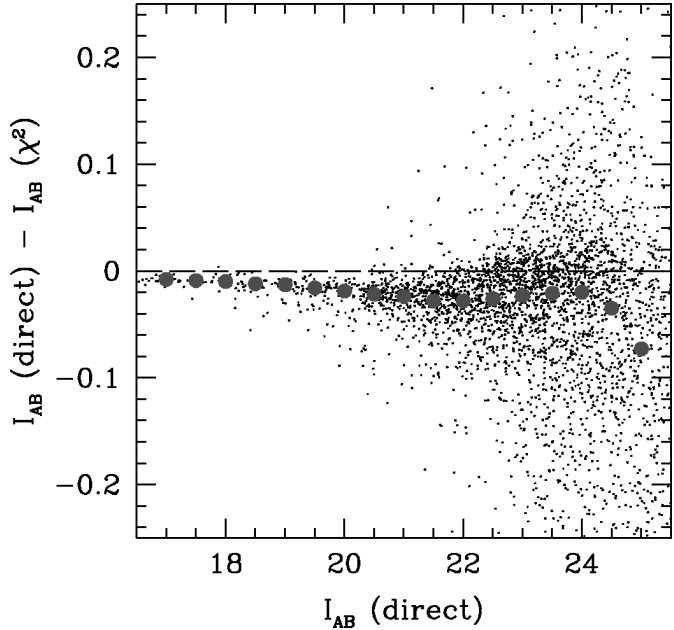}}
  \caption{The difference between Kron (1980) total magnitudes computed in the
    direct 03hr $I-$band image and using the $\chi^2$ technique
    ($\chi^2$ detection image combined with $I-$ band photometry image)
    as a function of total magnitude measured in the $I-$band direct
    image (for clarity only 1/4 of all points are shown). The filled
    shaded points line shows the median difference in half magnitude
    intervals.}
  \label{fig:compchis}
\end{figure}

As object parameters crucial to galaxy photometry, such as half-light
radius, are extracted from the detection image when using
\texttt{sextractor} in dual-image mode (in addition to the normal
($x$,$y$) centroids) we wished to ensure that the use of the $\chi^2$
image did not bias our derived (total) magnitudes. In
Fig.~\ref{fig:compchis} we plot the difference in galaxy total
magnitudes between the single band 03hr image and the dual-image mode
method ($\chi^2$ image and stacked image) as a function of total $AB$
magnitude in the single-band image. The filled shaded points show the
median magnitude difference in half-magnitude intervals. Until
$I_{AB}\sim24$, $I_{AB}~(direct)-I_{AB}(\chi^2) \lesssim 0.02$; for $24
< I_{AB} < 25.5$, $I_{AB}~(direct)-I_{AB}(\chi^2) \lesssim 0.1$. Beyond
$I_{AB}\sim24$ magnitudes computed using the direct image become
systematically brighter than the chisquared image, which is
most likely a consequence of the more reliable object profile
information contained in the chisquared image (which is comprised of
effectively a sum of object fluxes over all filters). In any case our
galaxy colour measurements, which use aperture magnitudes, are
unaffected by the application of this technique.

Our final catalogues consist of matched $V,I$ band catalogues for all
fields. For fields 14hr, 22hr and 03hr we have additional $B$ and $U$
band imaging. Magnitudes in our catalogues are Kron
\citep{1980ApJS...43..305K} ``total'' magnitudes computed using the
\texttt{sextractor} \texttt{$mag_{auto}$} parameter. We have also
carried out a comparison between these magnitudes and those computed
using the software employed in \citet{1986A&A...154...92L} and the
``Oxford'' galaxy photometry package described in
\citet{1991MNRAS.249..498M}.  We find no evidence of any systematic
differences between these three softwares. Throughout this paper we
measure colours in an aperture of $1.5''$ radius.

We also perform star-galaxy separation using the $r_h$ parameter from
\texttt{sextractor}, which is carried out on the $I-$ band catalogue.
This parameter measures the radius which encloses half the object flux.
Star-galaxy separation is not carried out faintwards of
$I_{AB}\sim21.5$; in any case for high galactic latitude fields like
ours galaxies outnumber stars by a large fraction at these faint
magnitudes \citep{1996AJ....112.1472R}.  The bright limit of our
catalogue (above which all galaxies and stars are saturated) is
$I_{AB}\sim18.5$.

\section{Galaxy counts and colours}
\label{sec:galaxy-counts}

\subsection{$I_{AB}-$ and $B_{AB}-$ selected galaxy counts}
\label{sec:i_ab--b_ab}

\begin{figure}
\resizebox{\hsize}{!}{\includegraphics{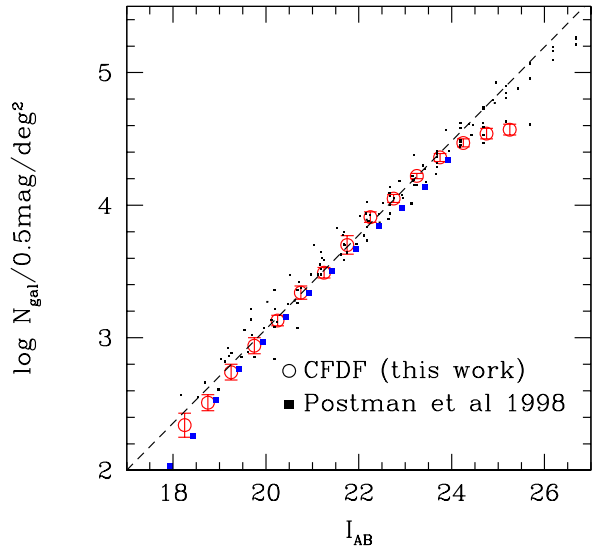}}
  \caption{$I_{AB}-$ band galaxy counts for the four
    fields in the CFDF survey (open circles). The dashed line shows a
    least-squares fit to the bright data ($22<I_{AB}<24$). The
    error-bar on each point corresponds to the field-to-field variance.
    The filled squares show galaxy counts from
    \citet{1998ApJ...506...33P}; the small points are from the
    compilation of Metcalfe et al.  (2000). }
  \label{fig:countsIAB}
\end{figure}

\begin{figure}
\resizebox{\hsize}{!}{\includegraphics{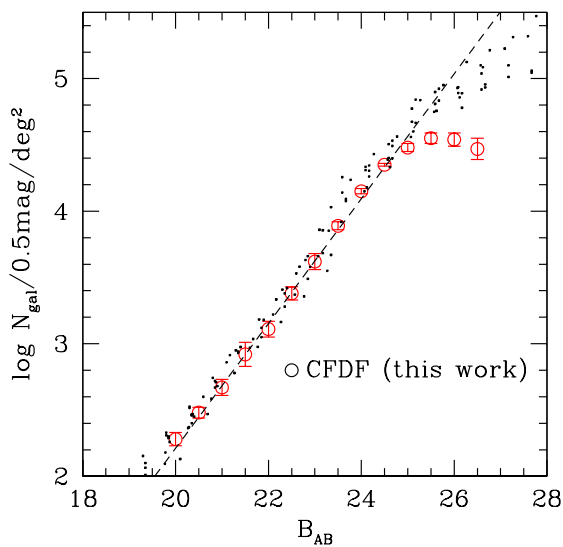}}
  \caption{As in Fig.~\ref{fig:countsIAB} but for the $B_{AB}$
    selected galaxy counts.}
  \label{fig:countsBAB}
\end{figure}

In Fig.~\ref{fig:countsIAB} and Fig.~\ref{fig:countsBAB} we present
$I-$ and $B-$band galaxy counts as a function of $AB$ magnitude derived
from the CFDF survey (open thick circles) compared with the compilation
of \citet{MSF}(small points).  For the $I-$ band we have four
independent fields, or approximately $\sim 10^5$ galaxies; for the $B-$
band there are currently three fields with a total of $\sim 0.75\times
10^5$ sources. In both these filters we find an excellent agreement
with the literature compilations extending over almost six orders of
magnitude.

We find the fitted slope, $\alpha$, for $20 < I_{AB} < 24$ to be
$0.35\pm0.02$, which agrees quite well with the value quoted in
\citet{MSF} of 0.33 for the slightly deeper limits of $21 < I_{kc} <
25$. Note that all our $I-$data, from $I_{AB}\sim18$ to $I_{AB}\sim24$
(beyond which the effects of incompleteness begin to be important) is
consistent with a constant slope in the number-magnitude relation.

We also compare our counts to those of \citet{1998ApJ...506...33P}
(filled squares). This dataset is a $16\deg^2$ survey comprising $256$
separate exposures of $900s$ each.  This dataset is $\sim 20\%$ below
the CFDF faintwards of $I_{AB}\sim22$ and remains so until the effects
of incompleteness begin to dominate the CFDF counts. Furthermore, in
our source counts we see no evidence of the inflection seen at
$I_{AB}\sim22$ in the \citeauthor{1998ApJ...506...33P} counts.

For the $B-$counts we find a fitted slope in the range $20 < B_{AB} <
24$ of $0.47 \pm 0.02$, consistent with the value of $0.50$ quoted in
\citet{1999ApJ...516..563B}.

\subsection{Galaxy colours}
\begin{figure}
\resizebox{\hsize}{!}{\includegraphics{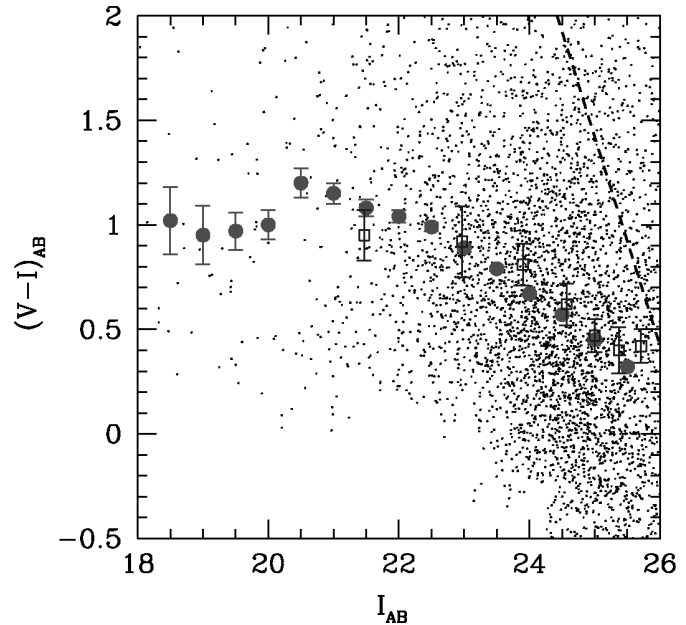}}
  \caption{Median $(V-I)_{AB}$ galaxy colour, measured as a function of
    $I_{AB}$ total magnitude, for galaxies in the CFDF-11hr field (for
    clarity only 1/8 of all galaxies are shown). The filled circles
    show the median galaxy colour in half magnitude intervals; the open
    squares show the median $(V-I)_{AB}$ colours measured in the NTT
    deep field \citep{1999A&A...341..641A}. The dashed line shows the
    nominal colour incompleteness, based on the detection limits given
    in Table~\ref{tab:cfdf.fields}. }
  \label{fig:VIcol}
\end{figure}

In Fig.~\ref{fig:VIcol} we show $(V-I)_{AB}$ colours measured as a
function of $I_{AB}$ total magnitude for galaxies in the CFDF 11hr
field (for clarity only 1/8 of all galaxies have been shown). The
dashed line represents the colour incompleteness computed using the
limiting magnitudes given in Table~\ref{tab:cfdf.fields}. Our median
$(V-I)_{AB}$ colours in half magnitude slices (filled circles) shows
good agreement with colours derived from the NTT deep field
\citep{1999A&A...341..641A}, shown as open squares, and this agreement
continues to the completeness limit of our data.

\label{sec:galaxy-colours}

\section{Measurements of $\omega(\theta)$ in the CFDF fields}
\label{sec:proj-corr-funct}

\subsection{Measuring $\omega(\theta)$}
\label{sec:measuring-omegatheta}

There is an extensive literature on the measurement of the projected
galaxy correlation function in deep imaging data
\citep{EBK,RSMF,BSM,1996ApJ...469..519H,1997ApJ...490...11W,2000MNRAS.318..913M}.
Here we will only briefly outline the relevant equations. We have
computed the two-point projected galaxy correlation function using the
\citet{LS} (LS) estimator,

\begin{equation}
\omega ( \theta) ={\mbox{DD} - 2\mbox{DR} + \mbox{RR}\over \mbox{RR}}, 
\label{eq:1.ls}
\end{equation}

with the $DD$, $DR$ and $RR$ terms referring to the number of
data-data, data-random and random-random galaxy pairs between $\theta$
and $\theta+\delta\theta$. In this work we use logarithmically spaced bins
with $\log(\delta\theta)=0.2$.

The fitted amplitudes quoted in this paper assume a power law slope for
the galaxy correlation function,
$\omega(\theta)=A_\omega\theta^{-\delta}$; however this amplitude must
be adjusted for the ``integral constraint'' correction, arising from
the need to estimate the mean density from the sample itself. This can
be estimated as \citep{RSMF},

\begin{equation}
C = {1 \over {\Omega^2}} \int \int \omega(\theta) d\Omega_1 d\Omega_2, 
\label{eq:6}
\end{equation}

where $\Omega$ is the area subtended by each of our survey fields. For
the CFDF fields, we find $C\sim 4A_\omega$ by numerical integration of
Equation~\ref{eq:5} over our field geometry and assuming that galaxies
closer than $1''$ cannot be distinguished. In deriving the correlation
amplitudes we then fit

\begin{equation}
\omega_{obs}(\theta) = A_\omega \theta^{-\delta} - C, 
\label{eq:5}
\end{equation}

where $\omega_{obs}(\theta)$ is our \emph{observed} measurement of
$\omega(\theta)$. (We note however that the integral constraint
correction only becomes important at larger scales ($\theta >
2\arcmin$) and that the large numbers of galaxies in our survey
combined with the large field of view of each patch means we can fit
for $A_\omega(1\arcmin)$ while neglecting the integral constraint,
providing the range of the fit is restricted to scales where the
integral constraint is not important). In our fitting procedure we use
the method of \citet{marquard} which takes into account that each
measurement of $\omega(\theta)$ at each angular separation is not
independent of the others.

To minimise computational requirements (necessary given the large
numbers of galaxies in our samples, typically 30,000 galaxies per
field) we use a sorted linked list method to compute the numbers of
galaxy pairs in each angular bin (and furthermore we have verified that
this method gives the same results as a direct-pair counting approach;
it is, however $\sim 2-3 \times$ faster). We use polygonal masks to
blank out regions surrounding bright stars, large galaxies, satellite
trails and cosmetic defects. We also discard data from CCD 8 because of
its poor charge transfer properties. Our masking strategy is quite
conservative, and (excluding CCD 8), around $15-20\%$ of the total area
is removed.  Because the 11hr and 22hr fields are composed of two
separate stacks which are rotated relative to each other, are able to
cover the entire $28'\times28'$ area for these fields.  We have
verified that the masks do not bias the determination of
$\omega(\theta)$ function by applying them to randomly generated
catalogues with the same number density as the real catalogues and
verifying that $\omega(\theta)$ is zero at all angular scales. We have
also tested the effect of masks on clustered catalogues generated using
the method of \citet{1978AJ.....83..845S}.  These tests (which were
carried out using the LS estimator) show that the application of masks
does not change our fitted correlation amplitudes. Note that we do not
apply a stellar dilution correction to our measured correlation
amplitudes as some authors do, as the form of the stellar counts for
high galactic latitude fields is not precisely known. However, this
correction is unlikely to raise the amplitude of the correlation
function by more than $10-15\%$, given that stars outnumber galaxies by
at least an order of magnitude beyond $I_{AB}\sim21.5$.

\subsection{$\omega(\theta)$ for $I_{AB}$ magnitude-limited samples}
\label{sec:magn-limit-sampl}

\begin{figure}
\resizebox{\hsize}{!}{\includegraphics{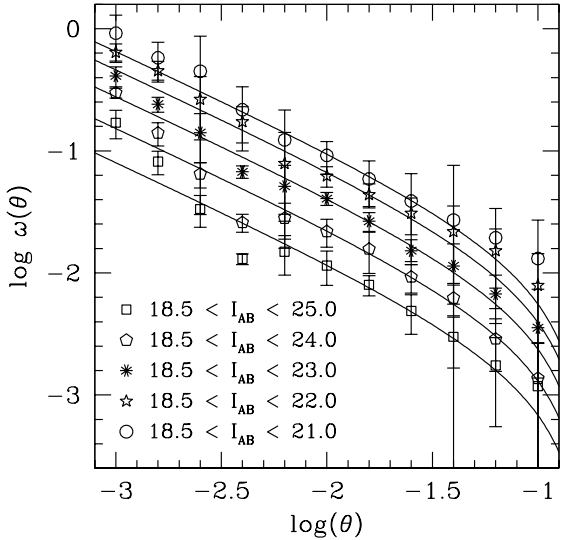}}
  \caption{The logarithm of the average amplitude of the angular correlation
    function, $\omega(\theta)$, as a function of the logarithm of the
    angular separation in degrees, for all fields in the CFDF.  The
    error bar on each bin is computed from the field-to-field variance.
    Each of the five different symbols shows a different
    magnitude-limited sample; the solid lines shows the fitted
    correlation amplitude. This weighted fit is carried out neglecting
    the three innermost bins and assuming a power-law slope of $ -0.8$
    for $\omega(\theta)$ and a value of $4.2$ for the integral
    constraint term $C$. Data from the 14hr field is not included in
    the faintest slice.}
  \label{fig:omega-raw}
\end{figure}

For each of our four fields, we divide our catalogue into magnitude
limited samples. Each sample reaches one half-magnitude deeper than the
previous, while the bright limit is kept at $I_{AB}=18.5$. (As the 14hr
field is shallower than the other fields, we do not measure
$\omega(\theta)$ on this field fainter than $I_{AB}=24.0$). These
samples are extracted from the $\chi^2$ catalogues prepared as outlined
in Sect.~\ref{sec:making-match-catal}; we have also performed the
same computation on single-band catalogues (i.e, those computed without
using an associated $\chi^2$ image for detection) and find identical
results.  Fig.~\ref{fig:omega-raw} shows the logarithm of the
amplitude of $\omega(\theta)$ averaged over the four fields of the CFDF
as a function of the logarithm of the angular separation in degrees,
for a number of magnitude-limited samples. In the range $-2.5< \log
(\theta) < -1.3$ ($0.2' <\theta < 3'$) our measurements follow the
expected power-law behaviour very well and at least until
$I_{AB}\sim24$ there is no evidence of significant excess power on
scales larger than the individual UH8K CCDs ($\sim 10'$), in agreement
with the evidence presented in Sect.~\ref{sec:syst-photm-errors-1}
that the systematic photometric errors in our fields are negligible. 

\nocite{2000astro.ph..7184C}
\nocite{1996MNRAS.279.1357L}

\begin{figure}
\resizebox{\hsize}{!}{\includegraphics{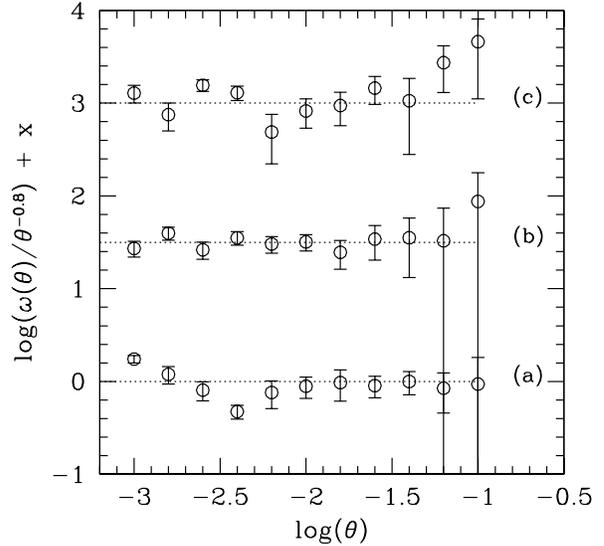}}
  \caption{Simulated measurements of $\omega(\theta)$, (b,c), compared
    with CFDF measurements for slice $18.5<I_{AB}<24.0$, (a). In each
    case the simulated measurements have been normalised by the fitted
    amplitude for the $18.5<I_{AB}<24.0$ slice, taking into account the
    integral constraint correction and assuming a $\delta=-0.8$ power
    law. The simulations (b,c) contain objects in the magnitude range
    $23.5<I_{AB}<24.5$ and $24.5<I_{AB}<25.5$ respectively. Each plot
    has been offset by an arbitrary amount $x$ where $x=0,1.5,3.0$. }
  \label{fig:omega-simulations}
\end{figure}

\begin{figure*}
\includegraphics[width=17cm]{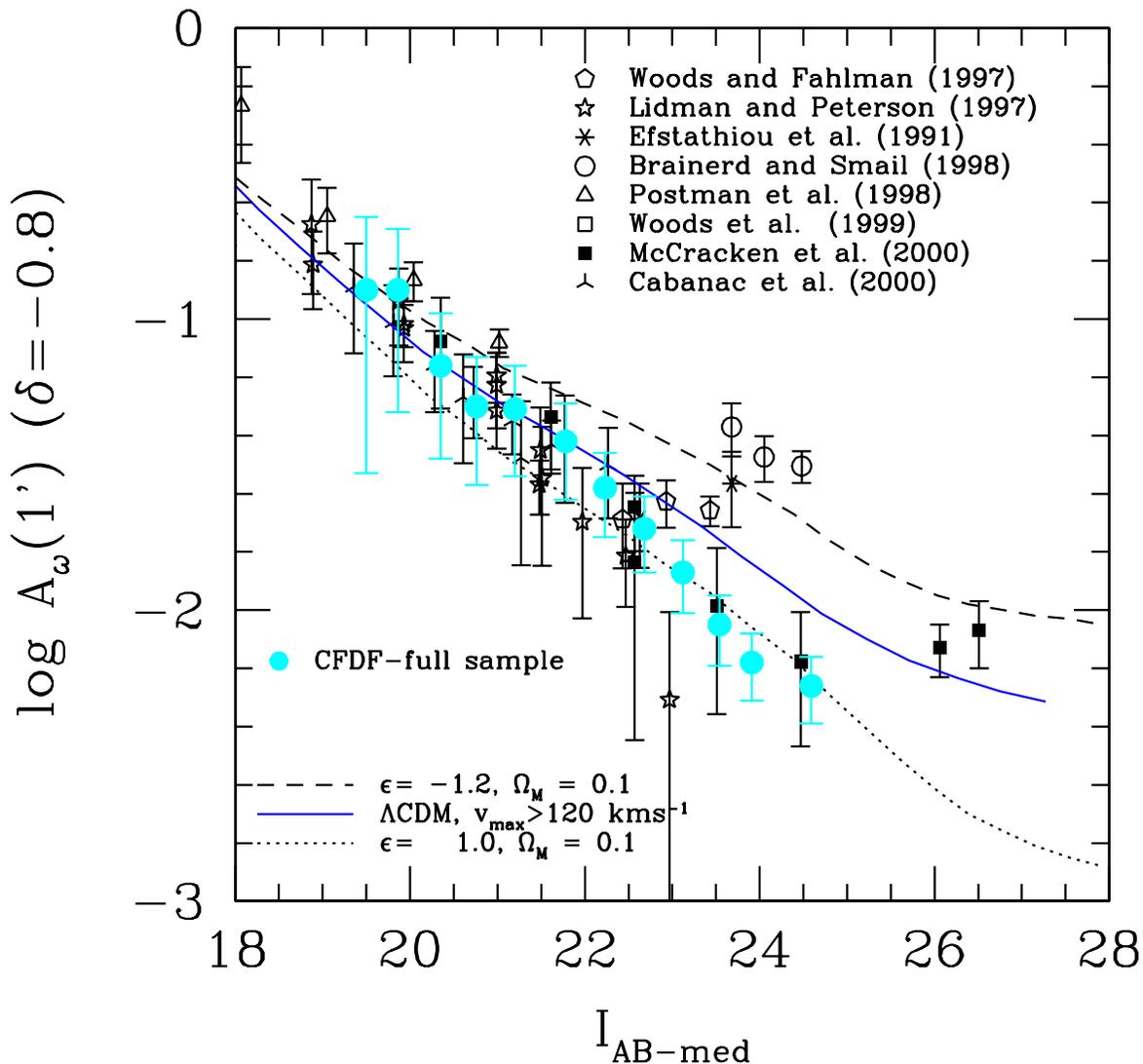}
  \caption{The logarithm of the fitted amplitude of the angular 
    correlation $\omega(\theta)$ at $1'$ as a function of sample median
    $I_{AB}$ magnitude (filled circles). Fits were computed assuming a
    power law of slope $\delta=-0.8$ for $\omega(\theta)$ and an
    integral constraint term calculated as described in the text. Error
    bars on the CFDF points are computed from the analytic expression
    of \cite{1994ApJ...424..569B}. The model curves are taken from
    McCracken et al. (2000) and show three different clustering
    evolution scenarios which are detailed in the text.}
  \label{fig:scaling1}
\end{figure*}

At $\theta < 0.2'$ we find at fainter magnitudes that $\omega(\theta)$
deviates from the expected power-law behaviour. We have attempted to
investigate the origin of this effect (which is seen in all our fields)
by carrying out an extensive set of simulations. In these simulations
we generate a catalogue with $\sim 40,000$ objects which have the same
correlation amplitude as the $I_{AB}\sim24$ sample.  Each of these
objects is assigned a random magnitude in a specified interval and then
added to the data frame. Object detection and photometry is carried out
in a box extracted at this location. A new catalogue is constructed
containing recovered total magnitudes for each object, and a magnitude
cut is then applied to this sample. Objects which are lost in this
process are typically those falling on or near bright stars or
galaxies, or those whose recovered total magnitude falls outside our
magnitude cut.  Masks are applied to this catalogue and
$\omega(\theta)$ computed using the same procedures used for the real
data. This procedure is then repeated for progressively fainter
magnitude slices.  (In constructing the simulated catalogue, two
simplifying assumptions were made, firstly that the input magnitude
distribution of objects is flat, and secondly we use objects with
Gaussian point-spread functions.)

In Fig.~\ref{fig:omega-simulations} we show the results of two of
these simulations, labelled (b) and (c).  The results displayed for (a)
show the measured correlations for the $18.5<I_{AB}<24.0$ magnitude
slice.  In order to display deviations from the power-law behaviour,
each slice has been normalised by the fitted amplitude for the
$18.5<I_{AB}<24.0$ slice, taking into account the integral constraint
correction and assuming a $\delta=-0.8$ power law. These simulations
cannot reproduce the depression in the correlation function found at
scales $\log(\theta)\sim -2.5$. We have also investigated if the
depression is caused by an excess of objects around bright stars, and
have found no evidence of such an excess. An important aspect of these
simulations, however, is that they confirm that excess power on large
scales only becomes important for the CFDF data for the faintest
magnitude ranges, $24.5<I_{AB}<25.5$.

In Fig.~\ref{fig:scaling1} we plot the fitted amplitude of
$\omega(\theta)$ at one arcminute, $A_\omega(1\arcmin)$, averaged over
all our fields, (filled circles) as a function of the sample median AB
magnitude ($I_{AB-med}$). We have estimated the error bars on
$A_\omega$, $\delta A_\omega$, using an analytic approximation
introduced by \cite{1994ApJ...424..569B} and further developed by
\cite{1996ApJ...470..131S}; see also Arnouts et al., in preparation.
for a detailed explanation of the application of this approximation.
The \citeauthor{1994ApJ...424..569B} estimator relies on a knowledge of
the higher-order moments of the galaxy correlation function, $S_3$ and
$S_4$; we have estimated these quantities directly from the CFDF
dataset and they will be presented in a future work (Colombi et al., in
preparation). At bright magnitudes ($I_{AB-med}\sim 18$), $\delta
A_\omega$ is dominated by the Poisson error component; however,
faintwards of $I_{AB-med}\sim 22$, the main component of $\delta
A_\omega$ in our field consists of cosmic variance (or ``finite
volume'') effects.  Our analytic estimates of this effect indicate that
the errors estimated empirically from the field-to-field variance of
our four fields may underestimate the total error at these magnitudes
by around $\sim 20 \%$.

Our fitted amplitudes were computed assuming a $-0.8$ slope for
$\omega(\theta)$ (to allow a comparison with the literature), and over
the range $0.2' <\theta < 3'$. We have also attempted to fit for
$A_\omega$ only at larger separations ($\sim 3\arcmin$), but our survey
fields are not large enough to detect any scale-dependence in the
$A_\omega-I_{AB-med}$ relation.  Additionally,
Fig.~\ref{fig:scaling1} shows a compilation of recent measurements
of $A_\omega$ from the literature.  These literature measurements all
assume a fixed slope of $\delta=-0.8$, with the exception of the
\citeauthor{1998ApJ...506...33P} survey, in which the slope was
allowed to vary with the fit. To allow a fair comparison with the
other authors, and with our work, the
\citeauthor{1998ApJ...506...33P} points are not plotted faintwards of
$I_{AB-med}\sim 21$, where their fitted slopes begin to differ
significantly from $-0.8$. 

\begin{figure}
\resizebox{\hsize}{!}{\includegraphics{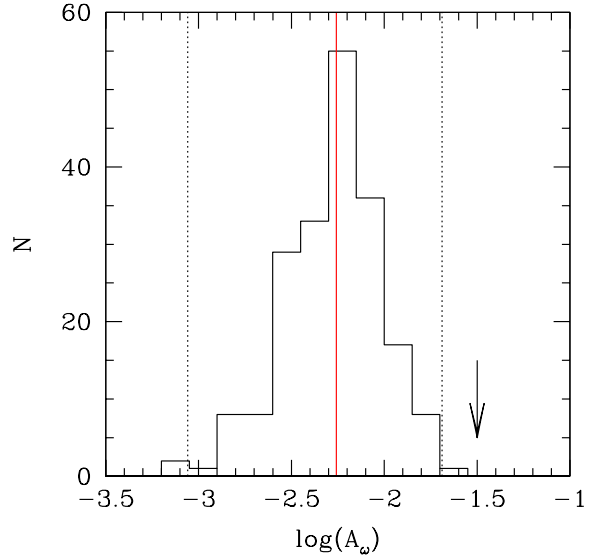}}
  \caption{The logarithm of the fitted correlation amplitude at
    $1\arcmin$, $\log(A_\omega)$ measured on 200 $7\arcmin \times
    7\arcmin$ sub-areas extracted from the 11hr and 22hr fields.
    Galaxies extracted have magnitudes in the range $18.5 < I_{AB} <
    25.5$. The dotted lines represent $\pm 3\sigma$ confidence limits
    about the mean value, shown by the solid line. The arrow shows
    approximately the $A_\omega(1\arcmin)$ measured by
    \citet{1998ApJ...494L.137B}.}
  \label{fig:variance}
\end{figure}

At bright magnitudes ($19 < I_{AB-med} < 22$) we find that our
observations are compatible with almost all the data from the
literature compilation. At fainter magnitude ranges, ($22 < I_{AB-med}
< 24$), our observations clearly favour a low amplitude for $A_\omega$.
At $I_{AB}\sim24 $ they are least a factor of ten below the value of
$A_\omega$ found by \citet{1998ApJ...494L.137B}. This work covers a
small area ($\sim 50$~arcmin$^2$) and consists of individual unconnected
pointings. As has been suggested before \citep{2000MNRAS.318..913M} the
discrepancy may be due to field-to-field variance in the galaxy
clustering signal.  To provide a more direct answer this question, we
extract 200 regions of $50$~arcmin$^2$ from the 22hr and 11hr fields
(these two have complete coverage as a consequence of bonette
rotations). On each of these sub-regions we measure $A_\omega$ as for
the full sample. In Fig.~\ref{fig:variance} we show the histogram of
the fitted values for the 11hr and 22hr fields; from this we estimate
that a $\pm 3\sigma$ error bar indicates a dispersion of $\times 10$ in
$A_\omega$.  It is clear that the error bars on
\citeauthor{1998ApJ...494L.137B} measurement underestimate the true
error by a large amount.

We have also determined $A_\omega(1\arcmin)$ in one-magnitude slices,
for instance, $19.5<I_{AB}<20.5$, $20.5<I_{AB}<21.5$ to the limit of
our survey, as an additional check. These measurements are considerably
more noisy than the integrated measurements presented above because of
the smaller numbers of galaxies in each slice.  However, we find that
the derived $I_{AB-med}-\log A_\omega(1\arcmin)$ relation is very
similar to what is presented in Fig.~\ref{fig:scaling1}.

\subsection{Dependence of slope on magnitude}
\label{sec:depend-slope-magn}

The large numbers of galaxies in our survey allows us to make a direct
measurement of $\delta$, the slope of $\omega(\theta)$ as a function of
sample limiting magnitude. Several attempts at this measurement have
been carried out but with generally inconclusive results.
\citet{1995ApJ...439...14N} find $\delta=-0.5$ at $g<25$ based on two
independent fields each the size of one of our CFDF fields. Shallower
wide-angle surveys like those of \citet{1999MNRAS.307..703R} and
\citet{2000astro.ph..7184C} find no evidence for deviation from a slope
of $\sim 0.8$. One difficulty in this measurement is that excess power
on large scales, such as can be produced by zero-point variations, can
produce an artificially shallow slope.  However, we have already
demonstrated that our magnitude zero-point errors are small across our
mosaics (Fig.~\ref{fig:syserrors1}) and that differential
incompleteness only becomes significant for measurements of
$\omega(\theta)$ in the CFDF fields at very faint magnitudes
(Fig.~\ref{fig:omega-simulations}).  Nevertheless, we adopt a
cautious approach and use two different methods to measure $\delta$.

\begin{figure}
\resizebox{\hsize}{!}{\includegraphics{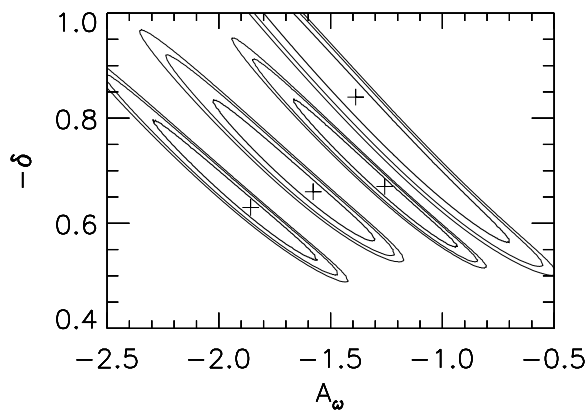}}
  \caption{$\chi^2$ contours showing best-fitting amplitudes and slopes
    (plus symbols) for the four faintest CFDF samples, from right to
    left, $18.5<I_{AB}<22,23,24,25$. The three contours show the
    $1\sigma$ (thick contour), $2\sigma$ and $3\sigma$ confidence
    levels.}
  \label{fig:slopes_chisq}
\end{figure}

\begin{figure}
\resizebox{\hsize}{!}{\includegraphics{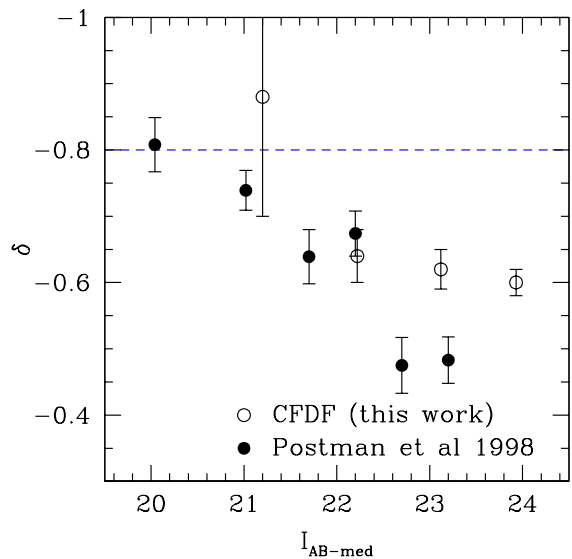}}
  \caption{The fitted slope of $\omega(\theta)$ as a function of sample 
    median magnitude (open circles) for galaxies with
    $18.5<I_{AB}<22,23,24,25$ averaged over the CFDF fields.  Points
    are also shown from Postman et al 1998 (filled circles).  Error
    bars on the CFDF points are computed from the variance over our
    four survey fields. }
  \label{fig:slopes}
\end{figure}

Firstly we compute $\chi^2$ contours from the average correlation per
angular bin over all the fields (Fig.~\ref{fig:omega-raw}), shown in
Fig.~\ref{fig:slopes_chisq}. In our second method we compute the slope
independently for each field and measure the field-to-field standard
deviation of this fit, which is shown in Fig.~\ref{fig:slopes},
together with a comparison with the fitted slopes from
\citet{1998ApJ...506...33P}, computed for a fitting range of
$0.5\arcmin<\theta<5\arcmin$.  An additional complication is that in
both cases our slope-fitting procedure requires an estimation of the
integral constraint (Equation~\ref{eq:5}), which in turn depends on the
slope. To limit these difficulties, we perform the fit only in the
range $ -0.2\arcmin < \theta< 1.9\arcmin$ where the effects of the
integral constraint are expected to be negligible.

Both methods indicate that at bright magnitudes, $18.5<I_{AB}<22.0$,
$\delta\sim -0.8$; at fainter magnitudes we detect a slight flattening
of the slope, with $\delta \sim -0.6$. The error bars in
Fig.~\ref{fig:slopes} show the error in $\delta$ for a given value of
$A_\omega$; from Fig.~\ref{fig:slopes_chisq} we see, however, that a slope
of $\delta=-0.8$ is still within $3\sigma$ of our best fit for all
magnitude ranges.

\subsection{Galaxy clustering as a function of colour}
\label{sec:colo-select-galaxy}

\begin{figure}
\resizebox{\hsize}{!}{\includegraphics{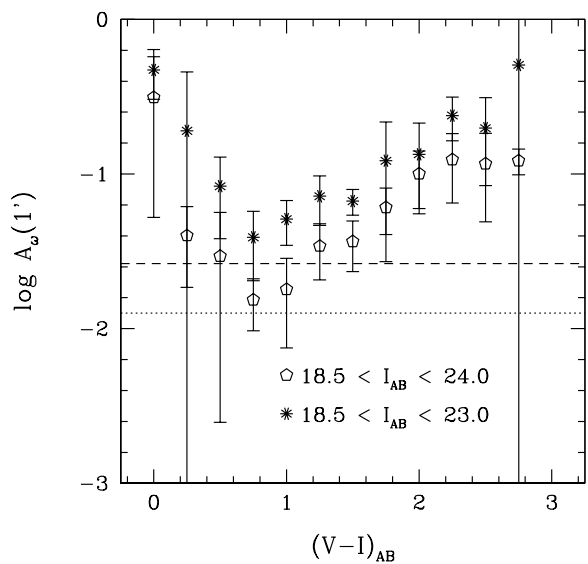}}
  \caption{The logarithm of the fitted amplitude of the angular 
    correlation $\omega(\theta)$ at $1'$ as a function of sample median
    $I_{AB}$ magnitude (as before, fits assume $\delta=-0.8$).
    $\log(A_\omega)(1\arcmin)$ was computed for 12 equally spaced bins
    of $0.25$ in $(V-I)_{AB}$. Error bars for each colour slice are
    computed from the field-to-field variance. Two magnitude ranges are
    shown, $18.5<I_{AB}<24.0$ (open pentagons) and $18.5<I_{AB}<23.0$
    (asterisks), together with the fitted amplitude for the full field
    samples, dotted and dashed lines respectively. }
  \label{fig:selgalcol}
\end{figure}

For \emph{all} galaxies in our sample we can measure $(V-I)_{AB}$
colours and we can use this to select subsamples by colour. In
Fig.~\ref{fig:selgalcol} we show $\log A_\omega (1\arcmin)$ as a
function of $(V-I)_{AB}$ colour for galaxies with $18.5<I_{AB}<24.0$
(open pentagons) and $18.5<I_{AB}<23.0$ (asterisks). For each of the
four fields, we divide our sample into twelve equally spaced bins in
colour each of width $0.25$ in $(V-I)_{AB}$. The error bars in $\log
A_\omega(1\arcmin)$ were computed from the field-to-field variance.
Also shown as the dotted and dashed lines is the amplitude of $\log
A_\omega(1\arcmin)$ for the full-field sample for these two slices.

We first note that both these cuts are relatively bright; at
$I_{AB}\sim24$ our catalogues are expected to be $\sim100 \%$ complete
(Figs.~\ref{fig:completeness},~\ref{fig:countsIAB}). Additionally, as
shown in Fig.~\ref{fig:VIcol}, the effect of colour incompleteness
should be minimal, although at $I_{AB}\sim24$ we may begin to lose some
extremely red objects from our sample. 

Our measurements clearly show that redder objects are more strongly
clustered, as has been widely reported for local populations
\citep{1995ApJ...442..457L}. We find that objects with
$(V-I)_{AB}\sim3$ have a clustering amplitude at least a factor of ten
higher than the full field population, shown as the dotted and
  dashed lines in Fig.~\ref{fig:selgalcol}. Interestingly, we also
find that the bluest objects in our survey, those with $(V-I)_{AB}\sim
0$, have a clustering amplitudes marginally higher than the full
sample. Furthermore, we find that \emph{none} of our colour selected
samples have clustering amplitudes below the full-field value. We
discuss the implications of these results in the following section.

\section{Discussion and comparison with model predictions}
\label{sec:disc-ompar-with}

\subsection{Modelling $A_\omega$}
\label{sec:model-predictions}

For $r<20h^{-1}$~Mpc, the spatial correlation function $\xi(r)$ can be
approximated by $\xi(r)=(r_0/r)^{\gamma}$, where, from the results of
local redshift surveys, $\gamma \sim -1.8$ and $r_0\sim 4.3h^{-1}$Mpc
\citep{1977ApJ...217..385G,1983ApJ...267..465D,1990MNRAS.247P...1M}.
One way to produce a prediction for the variation of $A_\omega$ with
sample limiting magnitude is to assume a functional form for the growth
of clustering $\xi(r,z)$, normally written as

\begin{equation}
 \xi(r,z)= h(z)\left({r_0\over r}\right)^\gamma ,
\label{eq:scal}
\end{equation}

where 

\begin{equation}
h(z)=(1+z)^{-(3+\epsilon)}.
\label{eq:hsca}
\end{equation}

This functional form is then integrated over redshift space using the
relativistic version of Limber's equation, 
\citep{1978MNRAS.182..673P,1977ApJ...217..385G,1953Apj...117..134},
assuming  that for $\theta \ll 1$ \citep{EBK}, 

\begin{equation}
\omega(\theta)= \sqrt \pi {\Gamma [(\gamma - 1) /2 ] \over
  {\Gamma(\gamma/2)}} {A \over {\theta ^{\gamma - 1} }} r_0^\gamma, 
\end{equation}
where $\Gamma$ is the incomplete gamma function, $\theta$ is the
angular separation and $A$ is given by
\begin{equation}
  A = \int^\infty_0 g(z)\left( dN \over {dz} \right)^2 dz / 
    \left[\int^\infty_0 \left( {dN \over {dz}} \right) dz \right]^2, 
\label{eq:aomega}
\end{equation}
with 
\begin{equation}
g(z)={h(z)\over{d_A^{\gamma-1}(z)}}\left ( {dr(z)\over{dz}}\right )^{-1},
\label{eqnew} 
\end{equation}
where $d_A(z)$ is the angular diameter distance and $dr(z)/dz$ is the
derivative of the proper distance. In this simple scenario, three cases
are of interest: clustering fixed in proper coordinates, in which
$\epsilon=0.0$; clustering fixed in co-moving coordinates which gives
$\epsilon=-1.2$.  Finally, the predictions of linear theory give
$\epsilon=0.8$.

Many papers have investigated the scaling of $A_\omega$ with magnitude
using the approach detailed above (see, for example, \citet{EBK}). To
interpret measurements of $A_\omega$, using these models, however,
involves making at least two critical assumptions: firstly, the form of
the redshift distribution $dN/dz$ for the faint galaxy population and
its evolution as a function of limiting magnitude; and secondly how
$\xi(r,z)$ scales with redshift (Equation~\ref{eq:hsca}). In the
following section we examine these two assumptions in turn. (Predicted
correlation amplitudes using this formalism are also sensitive to the
underlying cosmology, as the size of the volume element at a given
redshift is much lower for an Einstein-deSitter cosmology than for a
low-$\Omega$ universe.  However, to the median redshift of our survey
the difference in model predictions between open and flat-Lambda
cosmologies small. In this paper we assume that $\Omega_M\sim0.1$, in
agreement with recent observational evidence.)

\subsection{Validity of model assumptions}
\label{sec:valid-model-assumpt}

\begin{figure}
\resizebox{\hsize}{!}{\includegraphics{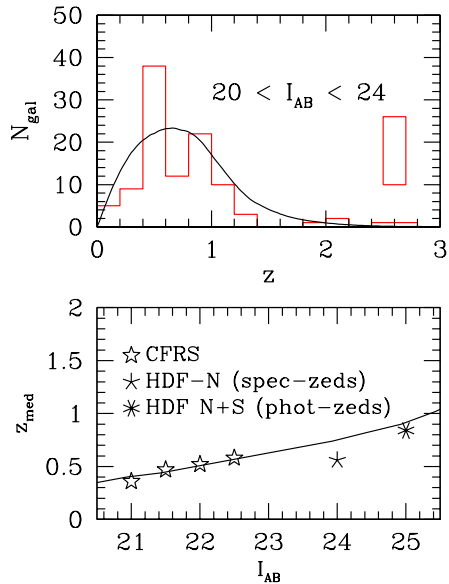}}
  \caption{Upper panel: the redshift distribution of faint galaxies 
    in the Hubble Deep Field North from the compilation of Cohen et al.
    (2000) (histogram), compared to the predictions of our model
    (smooth solid line). The open box represents the number of galaxies
    observed for which no redshift could be determined.  Lower panel:
    the relationship between median redshift $z_{med}$ predicted by our
    model (solid line) compared to the CFRS (five-pointed stars),
    spectroscopic redshifts in the HDF-N (cross) and photometric
    redshifts from the HDF N/S (asterisk). Poissonian error bars are
    smaller than the symbols in all cases and are not plotted.}
  \label{fig:dndz}
\end{figure}

Our $dN/dz$ is derived from luminosity evolution models which are
described fully in \citet{MSF}. Starting with the observed local galaxy
luminosity function and assuming a star-formation history for each
galaxy type these models are able to reproduce the observed numbers
counts, colours and redshift distributions of the faint galaxy
population to the limits of the current observations \citep{MSF}.
However, at $I_{AB}\sim24$ we may now directly test these model
redshift predictions against spectroscopic measurements made in the
Hubble Deep Field (HDF) \citep{2000ApJ...538...29C,WBD}.  In
Fig.~\ref{fig:dndz} (upper panel) we show the spectroscopic redshift
distribution for 120 galaxies in the HDF-North (with 16
non-detections represented as the open box) compared to the predictions
of our $\Omega_M=0.1,\Omega_\Lambda=0.0$ PLE model distribution (solid line).
In the lower panel we show the relationship between median redshift
$z_{med}$ and $I_{AB}$ limiting magnitude as predicted by this model
(note that we compute our median redshift in each case by considering
all galaxies brighter than the abscissa magnitude).  Additionally,
we show measurements from several other $I_{AB}-$ magnitude limited
redshift surveys, including the HDF sample used in the top panel. We
also show the median redshift derived from photometric redshifts for
two samples limited at $I_{AB}<25$ for the HDF N/S, kindly supplied to
us by S.  Arnouts.  The Poissonian error bars computed from all surveys
are smaller than the symbols and are not plotted. The \emph{true}
field-to-field variance may of course be much larger, but the fact that
we measure the same median redshift for both HDF-N/S catalogues suggest
that it is not.

Despite these qualifying remarks, we conclude that our luminosity
evolution models provide an acceptable fit to the observed redshift
distributions at least to the depth to which we measure galaxy
clustering in the CFDF photometric catalogues ($I_{AB-med}\sim24$, or
equivalently $I_{AB}\sim24.5$). They are able to reproduce both the
\emph{trend} of $z_{med}$ with $I_{AB}$ and the dispersion in redshift
at a given magnitude slice. Furthermore, as demonstrated in Metcalfe et
al., they also correctly predict the numbers of $2<z<3$ galaxies. We
therefore conclude that our modelling of $dN/dz$ is not a major source
of uncertainty in our prediction of $A_\omega$.

Our second assumption, that the growth of galaxy clustering can be
expressed as in Equation~(\ref{eq:hsca}), is more problematic.
Clustering measurements of Lyman-break galaxies
\citep{1998ApJ...505...18A,1998ApJ...503..543G}, have already indicated
that the ``epsilon'' formalism does not provide an acceptable fit to
the observations . Similar results have also been found for
measurements of $r_0(z)$ in the HDF-North \citep{1999MNRAS.310..540A}.
Can our clustering measurements in the CFDF be successfully matched by
this model?  In Fig.~\ref{fig:isca_new_plot} we show our measurements
of $A_\omega(1\arcmin$) compared to prediction of our models for
$\epsilon=-1.2,0.0,1.0,2.0$ (long dashed, solid, dotted and dashed
lines respectively), assuming $r_0=4.3h^{-1}$~Mpc, $\Omega_M=0.1,
\Omega_\Lambda=0.0$ and $\delta=-0.8$. (We note that clustering
predictions for an $\Omega-\Lambda$ cosmology are very similar to
zero-$\Lambda$ cosmology). For clarity we omit the literature
compilation shown previously. In the magnitude range
$18.5<I_{AB-med}<22$, we see that our observations are consistent with
$\epsilon \sim 0$. However, faintwards of $I_{AB}\sim22$, our observed
clustering amplitudes decline more rapidly than the model predictions.
By $23.0<I_{AB}< 24.0$ our observations are consistent with $\epsilon
\gtrsim 1$. From Fig.~\ref{fig:isca_new_plot} it is clear that the
$\epsilon \sim 0$ model cannot match simultaneously both bright and
faint observations in the range $18.5<I_{AB-med}<24.0$.  Furthermore,
rapid growth of clustering for the entire sample ($\epsilon\sim2$) is
marginally excluded because it produces correlations which are already
too low by $I_{AB-med}\sim 22$ to match our observations.  Furthermore,
allowing $r_0$ (i.e., $r_0(z=0)$) to vary merely changes the
normalisation at $I_{AB}\sim18.5$ (which is already in agreement with
our observations) but not the \emph{slope} of the
$A_\omega(1\arcmin)-I_{AB-med}$ relation.  In
Sect.~\ref{sec:bias-growth-galaxy} we investigate the reason for this
discrepancy in more detail.

\subsection{Colour selected galaxy clustering}
\label{sec:colo-select-clust}

In Fig.~\ref{fig:selgalcol} we clearly see the dependence of
$\log(A_\omega)$ on $(V-I)_{AB}$ colour. To interpret this result, we
first note that the dependence of $(V-I)_{AB}$ colour on redshift and
morphological type is well established, thanks to extensive
spectroscopic surveys
\citep{1999ApJ...518..533L,CSH2,1995ApJ...455...96C,1995ApJ...455...50L}.
In particular, \citet{2000astro.ph..8504W}, using a large spectroscopic
sample, demonstrated that objects with $(V-I)_{AB}\sim3$ are
predominately massive elliptical galaxies at $z\sim0.8$. Furthermore,
clustering amplitudes have recently been measured for objects selected
to have extremely red colours in optical-infrared bandpasses
\citep{2000A&A...361..535D}. These objects have clustering amplitudes
$\sim 10\times$ higher than the full field population. It is probable
that these objects are closely related to our $(V-I)_{AB}\sim3$ sample;
for galaxies with $(V-I)_{AB}>3$ at $I_{AB}<23.0$, we find
$\sim0.3$~N$_{gal}$~arcmin$^{-2}$.  In \citet{2000A&A...361..535D}
using a selection of $(R-K_s)>5.0$ and the slightly brighter limit of
$K_s<19.0$, they find a surface density of
$0.5$~N$_{gal}$~arcmin$^{-2}$. The difference between our full field
clustering amplitude at $18.0<I_{AB}<23$ and the clustering of objects
selected with $(V-I)_{AB}\sim3$ is approximately the same as the
difference found by \citeauthor{2000A&A...361..535D} between their
$K_s$-selected sample and those of their extremely red objects.

Intriguingly, at the blue end of our selection, $(V-I)_{AB}\sim0$, we
also find a higher clustering amplitude than the full field sample,
although the error bars are large due to field-to-field variations (at
the $\sim 0.1$ magnitude level) in galaxy colours and the small numbers
of objects involved. We have repeated our measurement of
$A_\omega(1\arcmin)$ using an integrated selection (i.e., considering
only objects redder or bluer than a specified colour cut) and find a
similar effect. There is some evidence for this effect in the
literature: working with photographic data, and considering objects in
a somewhat brighter blue selected magnitude cut, $20 < B_J < 23.5$,
\citet{1996ApJ...460...94L} also found an enhancement of at least $\sim
10$ for the clustering amplitudes of the bluest objects in ($U-R_F$).

Although Lyman-break galaxies are expected to be flat spectrum objects
and therefore have $(V-I)_{AB}$ colours of $\sim 0$ their surface
densities are not large enough to produce the effect seen in
Fig.~\ref{fig:selgalcol}. The most likely explanation of this result
is that these objects constitute a low-redshift population whose higher
correlation amplitudes are a consequence of the lack of projection
effects which dilute the measured $A_\omega$'s. Some evidence for this
can been seen in Fig. 5. of \citet{1995ApJ...455...96C}; all objects
with $(V-I)_{AB}\sim 0 $ are at $z<0.3$. We also note that our red and
blue samples have very low cross-correlation amplitudes, which supports
the notion that the objects in our survey with $(V-I)_{AB}\sim3$ and
$(V-I)_{AB}\sim 0$ are separate populations at different redshifts.

\subsection{Biasing and the growth of galaxy clustering}
\label{sec:bias-growth-galaxy}

\begin{figure}
\resizebox{\hsize}{!}{\includegraphics{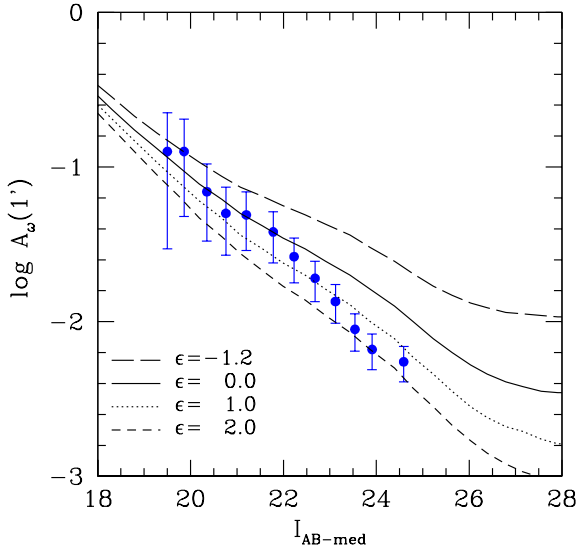}}
  \caption{The evolution of $\log(A_\omega(1\arcmin))$ 
    (assuming $\delta=-0.8, \Omega_M=0.1$ and $r_0=4.3h^{-1}$Mpc) for
    $\epsilon=-1.2,0.0,1.0,2.0$ (long dashed, solid, dotted, and dashed
    lines respectively). The filled points show the CFDF full sample.
    This plot is similar to Fig..~\ref{fig:scaling1} except the
    literature compilation has been omitted for clarity.}
  \label{fig:isca_new_plot}
\end{figure}

Given that the redshift distribution used in our models is in agreement
with our observations, then it is clear that the discrepancy evident in
Fig.~\ref{fig:isca_new_plot} between model predictions and
observations must be a result of evolution in the \emph{intrinsic}
properties of the galaxy population. Our simple model does not take
this into account. As a first step towards a more realistic description
of the data, we may try changing the form of the $r_0-z$ relation:
\citet{2000MNRAS.318..913M}, considered such a modification by adopting
the form of $\xi(r,z)$ derived from dark matter haloes with
$v_{max}>120$~kms$^{-1}$ identified in a large, high-resolution
$N-$body simulation \citep{1999ApJ...520..437K}. This is shown as the
solid line in Fig.~\ref{fig:scaling1}. However, because the form of
this relationship is very similar to the traditional ``epsilon-model''
in the range $0<z<1$, and because there are few $z>3$ galaxies in
samples limited at $I_{AB}<25$, the differences in predicted amplitudes
between this and the conventional formalism are small in the magnitude
ranges we consider.

The basic reason why the ``$\epsilon$''-models fail to reproduce the
clustering of Lyman-break galaxies and the observed form of $r_0(z)$ at
high redshift is that they implicitly ignore the existence of bias and
that how galaxies trace the underlying dark matter depends on the mass
of the dark matter halo
\citep{1984ApJ...284L...9K,1986ApJ...304...15B}. Measurements of galaxy
clustering in semi-analytic models, which provide a prescription for
how galaxies trace mass, show clearly that more luminous galaxies have
clustering amplitudes very different from less luminous ones
\citep{2001astro.ph..3092B,1999MNRAS.305L..21B,1999MNRAS.307..529K}.

Furthermore, it is now reasonably well established from observations
that locally the galaxy correlation length $r_0$ depends on
morphological type and colour
\citep{1997MNRAS.285L...5T,1995ApJ...442..457L,1976ApJ...208...13D},
and some evidence exists for a direct dependence between luminosity and
clustering amplitude \citep{1996ApJ...472..452B}. There are also
indications that these trends continue to higher redshifts
\citep{2000ApJ...542...57C,1996ApJ...461..534L}.  Furthermore, in our
dataset, the median field galaxy $(V-I)_{AB}$ colour changes by $\sim
0.4$ mag in the range $22<I_{AB}<24$ (Fig.~\ref{fig:VIcol}); from
Fig.~\ref{fig:selgalcol} we see that changes in colour of this
magnitude cannot produce the changes in amplitude of
$A_\omega(1\arcmin)$ seen in the data. For this reason we suggest that
the rapid decline in $A_\omega(1\arcmin)$ in the range $22< I_{AB-med}
< 24$ is a consequence of luminosity-dependent clustering segregation.

\begin{figure}
\resizebox{\hsize}{!}{\includegraphics{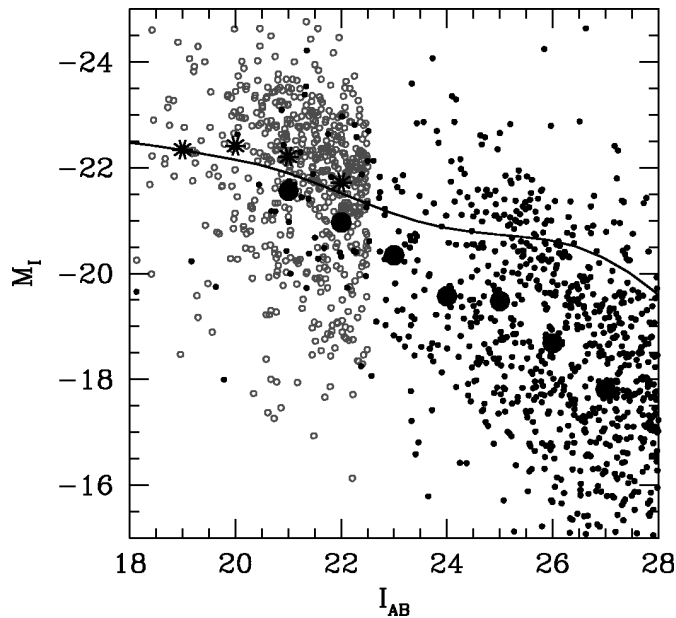}}
  \caption{Absolute $I_{AB}$ magnitudes ($M_I$) and median absolute 
    luminosities as a function of $I_{AB}$ magnitudes for galaxies
    selected from the CFRS (open circles, stars) and from the
    HDF-N (small and large filled circles).  Model predictions of our
    luminosity evolution model are shown as the solid line (In all
    cases luminosities are computed assuming $h=0.5$ and an open
    $\Omega_M=0.1$ cosmology).}
  \label{fig:lstar_plot}
\end{figure}

Extensive imaging and spectroscopic observations have demonstrated
that, for any magnitude limited sample, as we probe to fainter
magnitudes, the mean intrinsic luminosity of the field galaxy
population becomes progressively fainter. This is illustrated in
Fig.~\ref{fig:lstar_plot} where we show the absolute luminosity as a
function of apparent magnitude for galaxies in the CFRS survey
\citep{1995ApJ...455...50L} (open circles) and for galaxies in the
HDF-North computed using photometric redshifts from the photometric
catalogue of \citet{1999ApJ...513...34F}. For both these catalogues we
also show the median absolute magnitudes computed in half-magnitude
intervals of apparent magnitudes (open and filled circles for the CFRS
and HDF respectively). We also include an estimate of the median
differential luminosities $dL/dm$ from our luminosity evolution models
(solid line). We see that between $I_{AB}\sim20$ and $I_{AB}\sim22$,
within the CFRS sample, the median galaxy luminosity declines by $\sim
0.5$ magnitudes. In the range $22<I_{AB}<24$ we measure a decline of a
further magnitude, although the uncertainties in the absolute
luminosities computed from photometric redshifts in the HDF is probably
at least $\sim 0.5$ magnitudes. The decline in model luminosities seen
in the range $18.5 < I_{AB} < 22.5$ are a consequence of the steep
faint-end slope we adopt for the spiral galaxy luminosity function.
These faint galaxies at $I\sim22$ are predominately late-type galaxies,
as has been demonstrated by spectral and morphological classification
\citep{1998ApJ...499..112B,1998ApJ...496L..93D}.

\section{Conclusions}
\label{sec:conclusions}
In this paper we have introduced a new wide-field deep imaging survey,
the Canada-France deep fields project. Our survey covers a total area
of $\sim1$ deg$^2$ in four separate contiguous $0.25$~deg$^2$ fields.
We have demonstrated that our reduction procedures are robust. Our
internal astrometric errors are $\sim 0.06''$ and our systematic
photometric errors across each $0.25$~deg$^2$ field are $\lesssim 0.04$
magnitudes.

In each of our four fields our galaxy catalogue is $\sim 100\%$
complete for $18.5 < I_{AB} < 24.0$. We find that in this range the
$I_{AB}$ number-magnitude relation is well fitted by a line of
constant slope $d\log(N)/dm \sim 0.35\pm0.02$ in agreement with the
literature compilations. Our completeness drops to $50\%$ at
$I_{AB}\sim 25.5$.

Using this survey we have measured the projected-two point correlation
function $\omega(\theta)$ for a sample of 100,000 galaxies as a
function of sample median magnitude, $I_{AB-med}$, angular separation
$\theta$ and $(V-I)_{AB}$ colour to $I_{AB-med}\sim 25.0$. At
$I_{AB-med}\sim24$ we measure the amplitude of $\omega(\theta)$ at
$1\arcmin$, $A_\omega(1\arcmin)$ to an accuracy of $30\%$.  Our
conclusions are as follows:

1. We find that in the range $19.5 < I_{AB-med} < 24.0$, $A_\omega$
declines monotonically with sample limiting magnitude and that
throughout this range, $\omega(\theta)$ is well matched with a
power-law of slope $\delta$ for $0.2\arcmin < \theta < 3.0\arcmin$. At
bright magnitudes, $\delta\sim-0.8$; at $I_{AB-med}\sim23$, we find
$\delta\sim0.6$, although our observations are still compatible with
$\delta\sim-0.8$ at a $3 \sigma$ confidence level.

2. We find a clear dependence of $A_\omega(1\arcmin)$ on $(V-I)_{AB}$
colour for galaxies selected in two magnitude ranges, $18.5 < I_{AB} <
24.0$ and $18.5 < I_{AB} < 23.0$. Galaxies with $(V-I)_{AB}\sim3$ have
clustering amplitudes $\sim 10$ times higher than the full field
population. These objects are most probably evolved ellipticals at
$z\sim 1$. We also find some evidence (at the $\sim 1\sigma-2\sigma$
level) for slightly higher clustering amplitudes for the blue
$(V-I)_{AB}\sim0$ objects in our sample.

3. We discuss model predictions and current spectroscopic
determinations of the redshift distribution $dN/dz$ for the faint field
galaxy population. We conclude that for $19.5 < I_{AB-med} < 24.0$,
$dN/dz$ is now well determined. Using these predictions we find that
for low $\Omega$ cosmologies, and assuming a local galaxy correlation
length $r_0=4.3h^{-1}$Mpc, the growth of galaxy clustering,
parameterised by $\epsilon$, is consistent with $\epsilon\sim0$
  for galaxies in the magnitude range $19.5<I_{AB-med}<22.0$.

4. However, in the magnitude range $22.0<I_{AB-med}<24.0$, our
observations are consistent with $\epsilon \gtrsim 1$.  Models
  with $\epsilon \sim 0$ cannot match simultaneously measurements of
  $A_\omega(1\arcmin)$ at bright ($I_{AB-med}\sim19$) and faint
  ($I_{AB-med}\sim24$) magnitudes.

5. We demonstrate that one simple interpretation of this result is that
by $I_{AB-med}\sim24$ our sample is dominated by intrinsically faint
($M_I\sim-20$) galaxies which have considerably weaker correlation
lengths ($r_0\sim2-3h^{-1}$~Mpc) than the local galaxy population.

\begin{acknowledgements}
  HJMCC wishes to acknowledge the use of TERAPIX facilities at the
  Institut d'Astrophysique de Paris. We would also particularly like to
  thank Jean-Charles Cuillandre for assisting with the UH8K reductions,
  Frank Valdes at NOAO for patiently and thoroughly answering all our
  questions regarding MSCRED, and Emmanuel Bertin for many discussions
  concerning faint object photometry.  We wish also to thank Laurence
  Tresse for providing the CFRS $I-$ band absolute luminosities.
\end{acknowledgements}

\bibliographystyle{aa}

\end{document}